\newcommand{\figwidth}{0.475\textwidth}

\documentclass[prd,twocolumn,floatfix,amsmath,amssymb,aps,superscriptaddress,showpacs]{revtex4-1}

\pdfoutput=1

\usepackage{verbatim}
\usepackage[pdftex]{graphicx}
\usepackage{afterpage}
\usepackage{dcolumn}
\usepackage{bm}

\begin{document}

\thispagestyle{empty}


\title{Search for Higgs Bosons Produced in Association with $b$-Quarks}


\date{\today}

\affiliation{Institute of Physics, Academia Sinica, Taipei, Taiwan 11529, Republic of China} 
\affiliation{Argonne National Laboratory, Argonne, Illinois 60439, USA} 
\affiliation{University of Athens, 157 71 Athens, Greece} 
\affiliation{Institut de Fisica d'Altes Energies, ICREA, Universitat Autonoma de Barcelona, E-08193, Bellaterra (Barcelona), Spain} 
\affiliation{Baylor University, Waco, Texas 76798, USA} 
\affiliation{Istituto Nazionale di Fisica Nucleare Bologna, $^{aa}$University of Bologna, I-40127 Bologna, Italy} 
\affiliation{University of California, Davis, Davis, California 95616, USA} 
\affiliation{University of California, Los Angeles, Los Angeles, California 90024, USA} 
\affiliation{Instituto de Fisica de Cantabria, CSIC-University of Cantabria, 39005 Santander, Spain} 
\affiliation{Carnegie Mellon University, Pittsburgh, Pennsylvania 15213, USA} 
\affiliation{Enrico Fermi Institute, University of Chicago, Chicago, Illinois 60637, USA}
\affiliation{Comenius University, 842 48 Bratislava, Slovakia; Institute of Experimental Physics, 040 01 Kosice, Slovakia} 
\affiliation{Joint Institute for Nuclear Research, RU-141980 Dubna, Russia} 
\affiliation{Duke University, Durham, North Carolina 27708, USA} 
\affiliation{Fermi National Accelerator Laboratory, Batavia, Illinois 60510, USA} 
\affiliation{University of Florida, Gainesville, Florida 32611, USA} 
\affiliation{Laboratori Nazionali di Frascati, Istituto Nazionale di Fisica Nucleare, I-00044 Frascati, Italy} 
\affiliation{University of Geneva, CH-1211 Geneva 4, Switzerland} 
\affiliation{Glasgow University, Glasgow G12 8QQ, United Kingdom} 
\affiliation{Harvard University, Cambridge, Massachusetts 02138, USA} 
\affiliation{Division of High Energy Physics, Department of Physics, University of Helsinki and Helsinki Institute of Physics, FIN-00014, Helsinki, Finland} 
\affiliation{University of Illinois, Urbana, Illinois 61801, USA} 
\affiliation{The Johns Hopkins University, Baltimore, Maryland 21218, USA} 
\affiliation{Institut f\"{u}r Experimentelle Kernphysik, Karlsruhe Institute of Technology, D-76131 Karlsruhe, Germany} 
\affiliation{Center for High Energy Physics: Kyungpook National University, Daegu 702-701, Korea; Seoul National University, Seoul 151-742, Korea; Sungkyunkwan University, Suwon 440-746, Korea; Korea Institute of Science and Technology Information, Daejeon 305-806, Korea; Chonnam National University, Gwangju 500-757, Korea; Chonbuk National University, Jeonju 561-756, Korea} 
\affiliation{Ernest Orlando Lawrence Berkeley National Laboratory, Berkeley, California 94720, USA} 
\affiliation{University of Liverpool, Liverpool L69 7ZE, United Kingdom} 
\affiliation{University College London, London WC1E 6BT, United Kingdom} 
\affiliation{Centro de Investigaciones Energeticas Medioambientales y Tecnologicas, E-28040 Madrid, Spain} 
\affiliation{Massachusetts Institute of Technology, Cambridge, Massachusetts 02139, USA} 
\affiliation{Institute of Particle Physics: McGill University, Montr\'{e}al, Qu\'{e}bec, Canada H3A~2T8; Simon Fraser University, Burnaby, British Columbia, Canada V5A~1S6; University of Toronto, Toronto, Ontario, Canada M5S~1A7; and TRIUMF, Vancouver, British Columbia, Canada V6T~2A3} 
\affiliation{University of Michigan, Ann Arbor, Michigan 48109, USA} 
\affiliation{Michigan State University, East Lansing, Michigan 48824, USA}
\affiliation{Institution for Theoretical and Experimental Physics, ITEP, Moscow 117259, Russia}
\affiliation{University of New Mexico, Albuquerque, New Mexico 87131, USA} 
\affiliation{Northwestern University, Evanston, Illinois 60208, USA} 
\affiliation{The Ohio State University, Columbus, Ohio 43210, USA} 
\affiliation{Okayama University, Okayama 700-8530, Japan} 
\affiliation{Osaka City University, Osaka 588, Japan} 
\affiliation{University of Oxford, Oxford OX1 3RH, United Kingdom} 
\affiliation{Istituto Nazionale di Fisica Nucleare, Sezione di Padova-Trento, $^{bb}$University of Padova, I-35131 Padova, Italy} 
\affiliation{LPNHE, Universite Pierre et Marie Curie/IN2P3-CNRS, UMR7585, Paris, F-75252 France} 
\affiliation{University of Pennsylvania, Philadelphia, Pennsylvania 19104, USA}
\affiliation{Istituto Nazionale di Fisica Nucleare Pisa, $^{cc}$University of Pisa, $^{dd}$University of Siena and $^{ee}$Scuola Normale Superiore, I-56127 Pisa, Italy} 
\affiliation{University of Pittsburgh, Pittsburgh, Pennsylvania 15260, USA} 
\affiliation{Purdue University, West Lafayette, Indiana 47907, USA} 
\affiliation{University of Rochester, Rochester, New York 14627, USA} 
\affiliation{The Rockefeller University, New York, New York 10065, USA} 
\affiliation{Istituto Nazionale di Fisica Nucleare, Sezione di Roma 1, $^{ff}$Sapienza Universit\`{a} di Roma, I-00185 Roma, Italy} 

\affiliation{Rutgers University, Piscataway, New Jersey 08855, USA} 
\affiliation{Texas A\&M University, College Station, Texas 77843, USA} 
\affiliation{Istituto Nazionale di Fisica Nucleare Trieste/Udine, I-34100 Trieste, $^{gg}$University of Udine, I-33100 Udine, Italy} 
\affiliation{University of Tsukuba, Tsukuba, Ibaraki 305, Japan} 
\affiliation{Tufts University, Medford, Massachusetts 02155, USA} 
\affiliation{University of Virginia, Charlottesville, Virginia 22906, USA}
\affiliation{Waseda University, Tokyo 169, Japan} 
\affiliation{Wayne State University, Detroit, Michigan 48201, USA} 
\affiliation{University of Wisconsin, Madison, Wisconsin 53706, USA} 
\affiliation{Yale University, New Haven, Connecticut 06520, USA} 
\author{T.~Aaltonen}
\affiliation{Division of High Energy Physics, Department of Physics, University of Helsinki and Helsinki Institute of Physics, FIN-00014, Helsinki, Finland}
\author{B.~\'{A}lvarez~Gonz\'{a}lez$^w$}
\affiliation{Instituto de Fisica de Cantabria, CSIC-University of Cantabria, 39005 Santander, Spain}
\author{S.~Amerio}
\affiliation{Istituto Nazionale di Fisica Nucleare, Sezione di Padova-Trento, $^{bb}$University of Padova, I-35131 Padova, Italy} 

\author{D.~Amidei}
\affiliation{University of Michigan, Ann Arbor, Michigan 48109, USA}
\author{A.~Anastassov}
\affiliation{Northwestern University, Evanston, Illinois 60208, USA}
\author{A.~Annovi}
\affiliation{Laboratori Nazionali di Frascati, Istituto Nazionale di Fisica Nucleare, I-00044 Frascati, Italy}
\author{J.~Antos}
\affiliation{Comenius University, 842 48 Bratislava, Slovakia; Institute of Experimental Physics, 040 01 Kosice, Slovakia}
\author{G.~Apollinari}
\affiliation{Fermi National Accelerator Laboratory, Batavia, Illinois 60510, USA}
\author{J.A.~Appel}
\affiliation{Fermi National Accelerator Laboratory, Batavia, Illinois 60510, USA}
\author{A.~Apresyan}
\affiliation{Purdue University, West Lafayette, Indiana 47907, USA}
% ...
\collaboration{CDF Collaboration}
\thanks{With visitors from $^a$Istituto Nazionale di Fisica Nucleare, Sezione di Cagliari, 09042 Monserrato (Cagliari), Italy,
$^b$University of CA Irvine, Irvine, CA  92697, USA,
$^c$University of CA Santa Barbara, Santa Barbara, CA 93106, USA,
$^d$University of CA Santa Cruz, Santa Cruz, CA  95064, USA,
$^e$CERN,CH-1211 Geneva, Switzerland,
$^f$Cornell University, Ithaca, NY  14853, USA, 
$^g$University of Cyprus, Nicosia CY-1678, Cyprus, 
$^h$Office of Science, U.S. Department of Energy, Washington, DC 20585, USA,
$^i$University College Dublin, Dublin 4, Ireland,
$^j$University of Fukui, Fukui City, Fukui Prefecture, Japan 910-0017,
$^k$Universidad Iberoamericana, Mexico D.F., Mexico,
$^l$Iowa State University, Ames, IA  50011, USA,
$^m$University of Iowa, Iowa City, IA  52242, USA,
$^n$Kinki University, Higashi-Osaka City, Japan 577-8502,
$^o$Kansas State University, Manhattan, KS 66506, USA,
$^p$University of Manchester, Manchester M13 9PL, United Kingdom,
$^q$Queen Mary, University of London, London, E1 4NS, United Kingdom,
$^r$University of Melbourne, Victoria 3010, Australia,
$^s$Muons, Inc., Batavia, IL 60510, USA,
$^t$Nagasaki Institute of Applied Science, Nagasaki, Japan, 
$^u$National Research Nuclear University, Moscow, Russia,
$^v$University of Notre Dame, Notre Dame, IN 46556, USA,
$^w$Universidad de Oviedo, E-33007 Oviedo, Spain, 
$^x$Texas Tech University, Lubbock, TX  79609, USA,
$^y$Universidad Tecnica Federico Santa Maria, 110v Valparaiso, Chile,
$^z$Yarmouk University, Irbid 211-63, Jordan,
$^{hh}$On leave from J.~Stefan Institute, Ljubljana, Slovenia, 
}
\noaffiliation


\begin{abstract}

We present a search for neutral Higgs bosons $\phi$ decaying into
$b\bar{b}$, produced in association with $b$ quarks in $p\bar{p}$
collisions. This process could be observable in supersymmetric models
with high values of $\tan\beta$.  The event sample corresponds to
2.6~fb$^{-1}$ of integrated luminosity collected with the CDF~II
detector at the Fermilab Tevatron collider.  We search for an
enhancement in the mass of the two leading jets in events with three
jets identified as coming from $b$ quarks using a displaced vertex
algorithm.  A data-driven procedure is used to estimate the dijet mass
spectrum of the non-resonant multijet background.  The contributions
of backgrounds and a possible Higgs boson signal are determined by a
two-dimensional fit of the data, using the dijet mass together with an
additional variable which is sensitive to the flavor composition of
the three tagged jets.  We set mass-dependent limits on
$\sigma(p\bar{p}\rightarrow \phi b) \times \mathcal{B}(\phi\rightarrow
b\bar{b})$ which are applicable for a narrow scalar particle $\phi$
produced in association with $b$ quarks.
We also set limits on $\tan\beta$
in supersymmetric Higgs models including the effects of the Higgs boson
width.


\end{abstract}


\maketitle


\section{Introduction}
\label{s:intro}

The production of light Higgs bosons in association with $b$-quarks
can be significantly enhanced in the minimal supersymmetric standard
model (MSSM) or extensions thereof. This occurs when $\tan\beta$, the
ratio of the Higgs boson vacuum expectation values for up-type and
down-type quarks, is large. For $\tan\beta \sim 40$ the cross section
is expected to be a few picobarns~\cite{p:usladies}, giving a
production rate which could be observable in $p\bar{p}$
collisions at $\sqrt{s}=1.96$ TeV at the Fermilab Tevatron. In large
$\tan\beta$ scenarios the
pseudoscalar Higgs boson $A$ becomes degenerate with either the light
($h$) or heavy ($H$) scalar, doubling the cross section.

\begin{figure}
\caption{Neutral scalar production in association with $b$ quarks.}
\label{f:bg_to_bh}
\begin{center}
\begin{minipage}{0.135\textwidth}
\includegraphics[width=\textwidth]{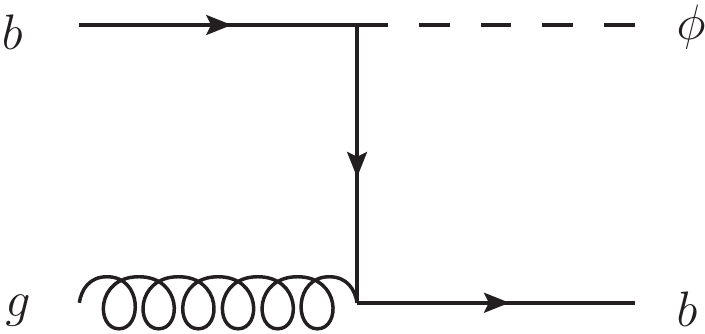}
\end{minipage}
\hspace{0.05\textwidth}
\begin{minipage}{0.135\textwidth}
\includegraphics[width=\textwidth]{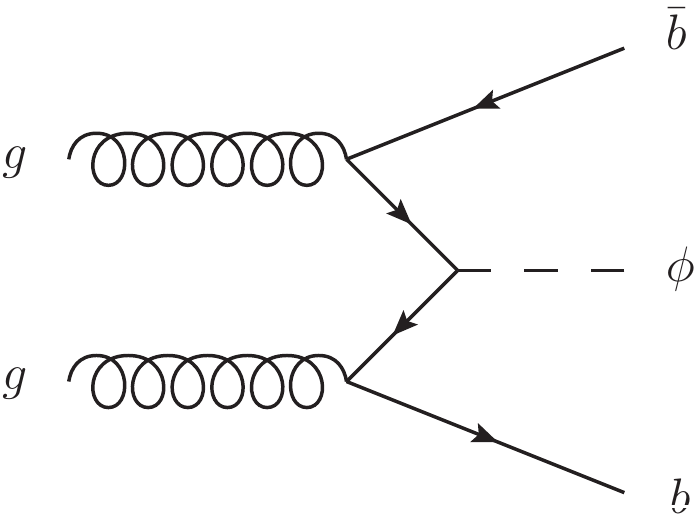}
\end{minipage}
\end{center}
\end{figure}

In the standard model (SM), the inclusive event yield of a light Higgs
boson in the $b\bar{b}$ decay channel is overwhelmed by strong
heavy-flavor pair production many orders of magnitude larger. For this
reason, searches for $H_{SM}\rightarrow b\bar{b}$ at the Tevatron rely
on associated production modes like $WH_{SM}$ and $ZH_{SM}$ where
backgrounds are restricted to those also containing a $W$ or $Z$. In
this paper we report on a search for $\phi\rightarrow b\bar{b}$, where
$\phi$ represents a narrow scalar such as $H_{SM}$ or the MSSM
Higgs bosons $h/H/A$, with the associated production $b\phi$ likewise
reducing the large heavy flavor backgrounds.  The production process
is illustrated in Fig.~\ref{f:bg_to_bh}.  Results for the $b\phi$
process in the case of Higgs boson decays to $b\bar{b}$ have been
previously obtained by D0~\cite{p:d04b-prl,p:d04b-1fb,p:d04b-5fb}, and
for inclusive or $b$-associated Higgs boson production in the
$\tau\tau$ decay mode by CDF~\cite{p:cdfditau-18fb},
D0~\cite{p:d0ditau-1fb,p:d0bditau-3fb}, and CMS~\cite{p:cmsditau}.  

\begin{figure}
\caption{The reconstructed $m_{12}$ distribution for simulated events
containing a 150~GeV/$c^2$ SM Higgs boson, for all events passing the
selection criteria and for only those where the two leading jets represent the
$b$ quarks from the Higgs boson decay (70\% of events for this mass).
No backgrounds are included.}
\label{f:higgs_match}
\begin{center}
\includegraphics[width=\figwidth]{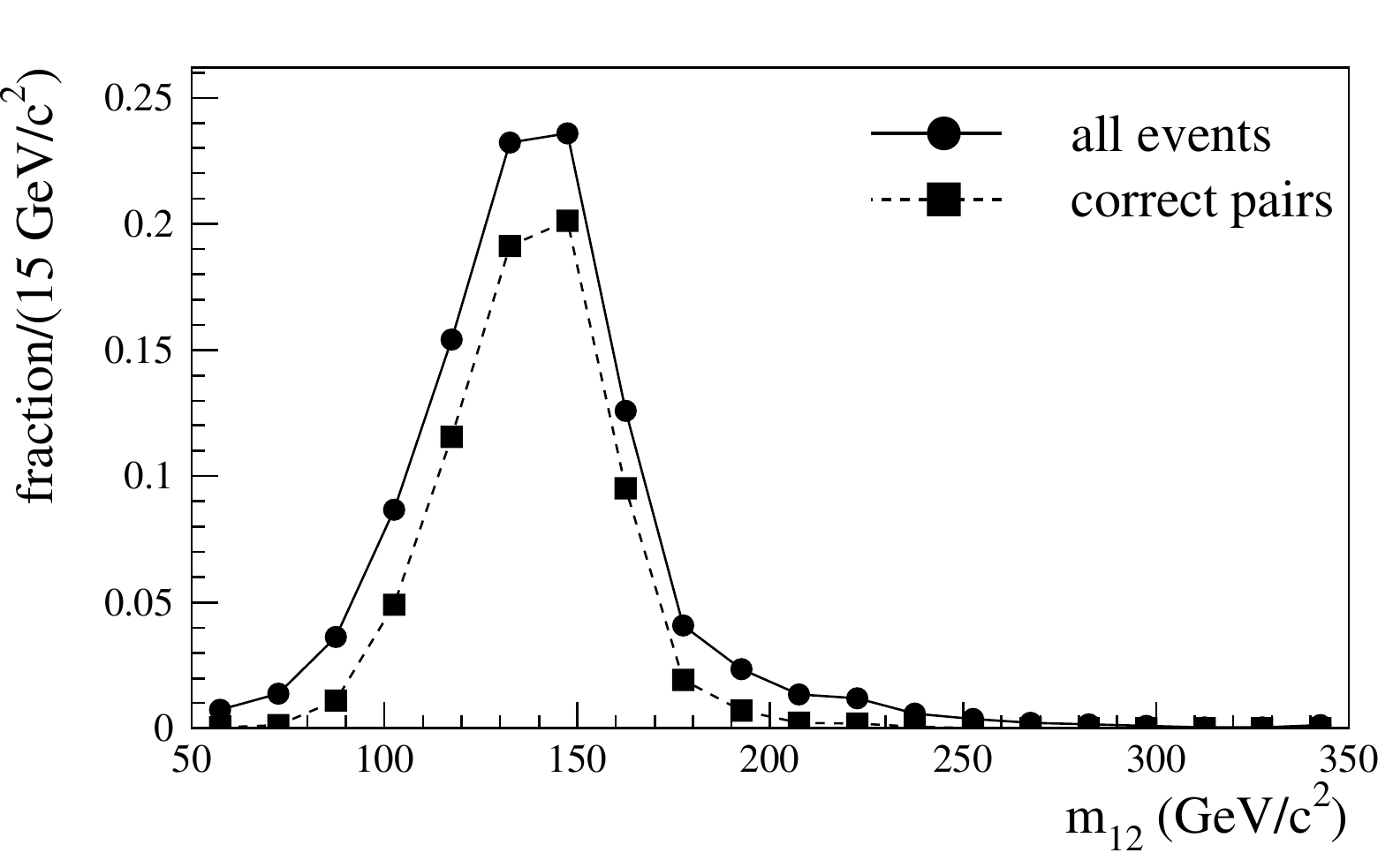}
\end{center}
\end{figure}

We search for resonance decays into $b\bar{b}$ in events containing
at least three $b$-jet candidates identified by displaced vertices
(``tagged'' hereafter). As the jets resulting from the resonance
decay are usually the most energetic jets in the event, we study the
invariant mass of the two leading jets in $E_T$, denoted
$m_{12}$. A signal would appear as an enhancement in the
$m_{12}$ spectrum.  An example $m_{12}$ distribution is shown in
Fig.~\ref{f:higgs_match}.

The background is predominantly QCD multijet production containing multiple
bottom or charm quarks. Events with single pairs of heavy flavor
also enter the sample when a third jet from a light quark or gluon is
mistakenly tagged.
We do not have precise {\em a priori} knowledge of the background
composition and kinematics, nor do we wish to rely upon a Monte Carlo
generator to reproduce it well~\cite{p:lannon,p:vistasleuth,p:paolo}. We have
instead developed a technique to model the $m_{12}$ spectrum for the
background in the triple-tagged sample in a data-driven manner,
starting from double-tagged events.

To enhance the separation between the flavor-dependent background
components and the possible resonance signal, we introduce a second
quantity $x_{tags}$, constructed from the invariant masses of the
secondary vertices constructed from the charged particle tracks in
each jet, which is sensitive to the flavor composition: three bottom
quark jets vs. two bottom quarks and one charm quark, etc. The
kinematic information in $m_{12}$ is then complemented by flavor
information in $x_{tags}$.

With data-driven estimates of the distributions of $m_{12}$ and
$x_{tags}$ for the backgrounds and Monte Carlo models for the neutral
scalar signal, we perform maximum likelihood fits of the
two-dimensional distribution of $x_{tags}$ versus $m_{12}$ in the data
to test for the presence of resonances in the triple-tagged
sample.  These fits are used to set limits on the cross section times
branching ratio for $\sigma(p\bar{p}\rightarrow b\phi) \times
\mathcal{B}(\phi\rightarrow b\bar{b})$ and on $\tan\beta$ in MSSM
scenarios.  Although the procedure has been optimized for the case of
production of a single resonance with the decay products predominantly
represented by the two leading jets in the event, the results can also
be interpreted in models of new physics with similar final states such
as pair production of color octet scalars~\cite{p:bogdan,p:bogdan2,p:aaron}.

In Sec.~\ref{s:detector} we briefly describe the CDF~II detector
subsystems upon which this analysis relies.  We discuss the data
sample and event selection requirements in Sec.~\ref{s:data}.  A
description of the signal simulation used for the search
is found in Sec.~\ref{s:signal}.  The data-driven background model is
presented in Sec.~\ref{s:bkgd}.  The systematic uncertainties on the
signal and background estimates are discussed in
Sec.~\ref{s:systematics}.  The results for the standard model and MSSM
interpretations are shown in Sec.~\ref{s:results}.  In
Sec.~\ref{s:conclusion} we summarize and conclude.

\section{The CDF II Detector}
\label{s:detector}

The CDF~II detector is an azimuthally and forward-backward symmetric
apparatus designed to study $p\bar{p}$ collisions at the Fermilab
Tevatron.  Details of its design and performance are described
elsewhere~\cite{p:cdfdet}, here we briefly discuss the detector
components which are relevant for this analysis.  The event kinematics
are described using a cylindrical coordinate system in which $\phi$ is
the azimuthal angle, $\theta$ is the polar angle with respect to the
proton beam, $r$ is the distance from the the nominal beam line, and
positive $z$ corresponds to the proton beam direction, with the origin
at the center of the detector.  The transverse $r-\phi$ (or $x-y$)
plane is the plane perpendicular to the $z$ axis.  The pseudorapidity
$\eta$ is defined as $-\ln (\tan (\theta/2))$.  The transverse
momentum of a particle is defined as $p_T = p\sin\theta$ and the
transverse energy as $E_T = E\sin\theta$.

A magnetic spectrometer consisting of tracking devices inside a 3-m
diameter, 5-m long superconducting solenoidal magnet with an axial
magnetic field of 1.4 T measures the momenta and trajectories of
charged particles.  A set of silicon microstrip detectors (L00, SVX,
and ISL)~\cite{p:layer00,*p:svx,*p:isl} reconstructs charged particle
trajectories in the radial range 1.5-28 cm, with a resolution on the
particle position at its closest approach to the beamline of 40~$\mu$m
in the transverse plane (including a 30~$\mu$m contribution from the
size of the beam spot).  A 3.1-m long open-cell drift chamber
(COT)~\cite{p:cot} occupies the radial range 40-137 cm.  Eight
superlayers of drift cells with 12 sense wires each, arranged in an
alternating axial and $\pm 2^\circ$ pattern, provide up to 96
measurements of the track position.  Full radial coverage of the COT
extends up to $|\eta|<1$ and of the silicon detectors up to
$|\eta|<2$.

A sampling calorimeter system arranged in a projective-tower geometry
surrounds the magnetic solenoid and covers the region up to
$|\eta|<3.6$.  The calorimeter is sectioned radially into 
lead-scintillator electromagnetic~\cite{p:cem,*p:pem} and iron-scintillator
hadronic~\cite{p:cha} compartments.  The central part of the
calorimeter ($|\eta|<1.1$) is segmented in towers spanning 0.1 in
$\eta$ and 15$^\circ$ in $\phi$.  The forward regions ($1.1 < \eta <
3.6$) are segmented in towers spanning 0.1 to 0.64 in $\eta$,
corresponding to a nearly constant 2.7$^\circ$ in $\theta$.  The
$\phi$ segmentation of the forward regions is 7.5$^\circ$ for $1.1 <
|\eta| < 2.11$ and 15$^\circ$ for $|\eta|>2.11$.

Drift chambers located outside the central hadronic calorimeters and
behind a 60 cm thick iron shield detect muons with
$|\eta|<0.6$~\cite{p:cmu}.
Gas Cherenkov counters with a coverage of $3.7 < |\eta| < 4.7$ measure
the average number of inelastic $p\bar{p}$ collisions per beam
crossing and thereby determine the luminosity~\cite{p:clc}.

\section{Data Sample and Event Selection}
\label{s:data}

This analysis is based on a data sample corresponding to an integrated
luminosity of 2.6~fb$^{-1}$ collected with the CDF~II detector between
February 2002 and July 2008.  The data are collected using a
three-level trigger system.  The first level requires two towers in
the central calorimeter with $E_T>5$~GeV and two tracks with
$p_T>2$~GeV/$c$ reconstructed in the COT.  The second level requires
two energy clusters in the calorimeter with $E_T>15$~GeV and
$|\eta|<1.5$~\cite{p:l2cal}, along with two tracks with
$p_T>2$~GeV/$c$ and impact parameter $|d_0|>100$~$\mu$m,
characteristic of heavy flavor hadron decays, reconstructed using the
level~2 silicon vertex trigger (SVT)
system~\cite{p:svt,*p:svtupgrade}.  The third level confirms the
level~2 silicon tracks and calorimeter clusters using a variant of the
offline reconstruction.  No matching is required between the tracks in
the silicon tracker and the calorimeter towers or clusters in the
trigger system.  

Due to the increasing Tevatron instantaneous
luminosity profile, a higher-purity replacement for this trigger was
implemented in July 2008 to stay within the constraints imposed by the
CDF data acquisition system.  Because the analysis is so
tightly coupled to the trigger requirements, analysis of the data
collected after July 2008 will require a separate dedicated study.

The offline selection requires at least three jets with $E_T>20$~GeV and
detector rapidity $|\eta|<2$.  The jets are reconstructed using a cone
algorithm with radius $\Delta R =
\sqrt{\Delta\phi^2+\Delta\eta^2}<0.7$, and are corrected for
calorimeter response and multiple interactions so that the energy
scale reflects the total $p_T$ of all particles within the jet cone.
In addition, only jets containing at least two tracks within a cone of
$\Delta R = 0.4$ around the jet axis satisfying the quality
requirements of the displaced vertex-finding algorithm
SECVTX~\cite{p:secvtx} are considered.  If more than three jets in the
event satisfy these requirements we consider up to the fourth leading
jet in the event selection requirements (see below).  Additional jets
satisfying the requirements beyond the fourth leading jet are allowed
but not used in the event selection.  No veto is applied for
additional jets not satisfying these cuts, but they are ignored when
we order the jets by $E_T$ for the purpose of identifying the leading
jets in the event.  At least two of the three or four jets which are
used for the event selection must match the positions of the
calorimeter clusters found by the second and third levels of the
trigger in $\eta$ and $\phi$.

The signal sample for this search is defined by requiring that the two
leading jets in the event and either the third or fourth leading jet
be tagged as $b$-jet candidates using SECVTX.  The two leading jets in
the event must also match the displaced tracks required by the level~2
trigger selection.  The track matching allows for the case where both
tracks are matched to either of the two leading jets, or where each of
the two leading jets has one of the tracks matched.  The matched track
requirements bias the properties of the displaced vertices found by
the SECVTX algorithm.  Restricting the track matching to only the two
leading jets simplifies the accounting of these biases at a cost of
20-25\% in efficiency relative to allowing the tracks to match any of
the SECVTX-tagged jets in the event.

We also select a
superset of the triple-tagged signal region by requiring both of
the two leading jets, or at least one of the two leading jets and
either the third or fourth leading jet, to pass the SECVTX tag and
level~2 track matching requirements.  This double-tagged sample is the
starting point for the background estimation procedure described in
Sec.~\ref{s:bkgd}.

We find 11 490 events passing the triple-tagged signal sample
requirements.  The double-tagged sample with both of the two leading
jets tagged contains 267 833 events, and the sample with at least one of
the two leading jets tagged and either the third or fourth jet tagged
contains 424 565 events.

\section{Signal Model}
\label{s:signal}

To compute the efficiency of this selection for neutral scalar
signal events, the cross section of the process being searched for must be
precisely defined.  We use the {\sc mcfm} program~\cite{p:mcfm3b} to
calculate the cross section for $bg \rightarrow H_{SM}+b_{jet}$ in the
standard model.  From this baseline the Higgs boson production rates
in supersymmetric models are obtained by scaling the
couplings~\cite{p:balazs,p:scenarios}.
If there is a gluon in the final state along with the outgoing $b$
quark ({\sc mcfm} does not simulate the Higgs boson decay) and they
are within $\Delta R < 0.4$ of each other, {\sc mcfm} will combine
them into a ``$b_{jet}$''; otherwise the $b$ quark alone serves as the
jet.  This $b_{jet}$ is the object upon which the kinematic cuts can
be applied.

\begin{figure}
\caption{Cross section for $bg \rightarrow H_{SM}+b_{jet}$ in the standard 
model calculated with {\sc mcfm}.}
\label{f:mcfm-xs}
\begin{center}
\includegraphics[width=\figwidth]{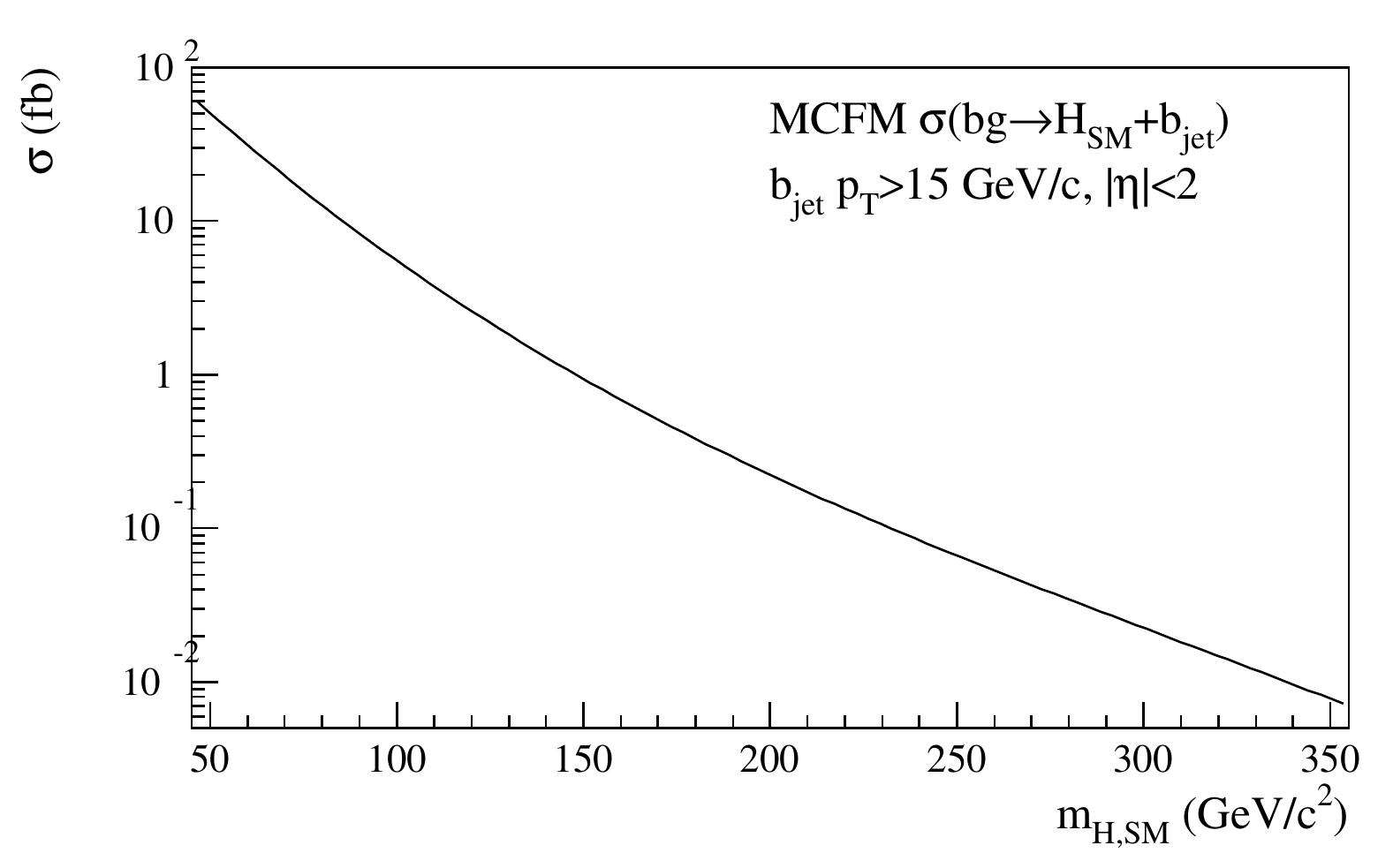}
\end{center}
\end{figure}

We calculate the cross section for $H_{SM}+b_{jet}$ in the SM, requiring
$p_T > 15$~GeV/$c$ and $|\eta|<2$ for the $b_{jet}$ to match the
acceptance of the SECVTX algorithm.  We use CTEQ6.5M~\cite{p:cteq65m}
parton distribution functions and set the renormalization and
factorization scales to $\mu_R = \mu_F = (2m_b+m_H)/4$ as suggested in
Refs.~\cite{p:bbh-scale,p:bbh-review}.  The cross section obtained as
a function of $m_{H,SM}$ is shown in Fig.~\ref{f:mcfm-xs}.  Cross
sections at the level of a femtobarn are not discernable in this final
state at the Tevatron, so in the SM this process is of little
interest.  In the MSSM, however, simple tree-level scaling of the
couplings and the degeneracy of the pseudoscalar $A$ with one of the
scalars $h/H$ enhances this cross section by a factor of
$2\tan^2\beta$.  For $\tan\beta = 50$ we therefore expect cross
sections of picobarns or more at the Tevatron.

The efficiency of the triple-tagged selection in events where the
neutral scalar decays into a $b\bar{b}$ pair is determined from
simulated data generated using the {\sc
pythia}~\cite{p:pythia1,*p:pythia2} Monte Carlo program and a full
simulation of the CDF~II detector~\cite{p:cdfsim}.  We generate
associated production of narrow scalars (specifically, SM Higgs
bosons) with additional $b$ quarks, and compare the kinematics of the
events to the momentum and rapidity distributions predicted by the
{\sc mcfm} calculation.  We find that the associated $b$ jets (those
not resulting from a Higgs boson decay) produced by {\sc pythia} are
more central than is predicted by {\sc mcfm}, while the other event
kinematics are in good agreement.  We correct the {\sc pythia} samples
to match the {\sc mcfm} predictions by reweighting the events based on
the pseudorapidity of the associated $b$ jets.  Further corrections
are applied in order to match the efficiencies of the SECVTX algorithm
and level~2 silicon tracking requirements to those measured in the CDF
data~\cite{p:secvtx}.

\begin{figure}
\caption{Selection efficiency for $b\phi$ events as a function of the
neutral scalar mass $m_\phi$.}
\label{f:acc_vs_mh}
\begin{center}
\includegraphics[width=\figwidth]{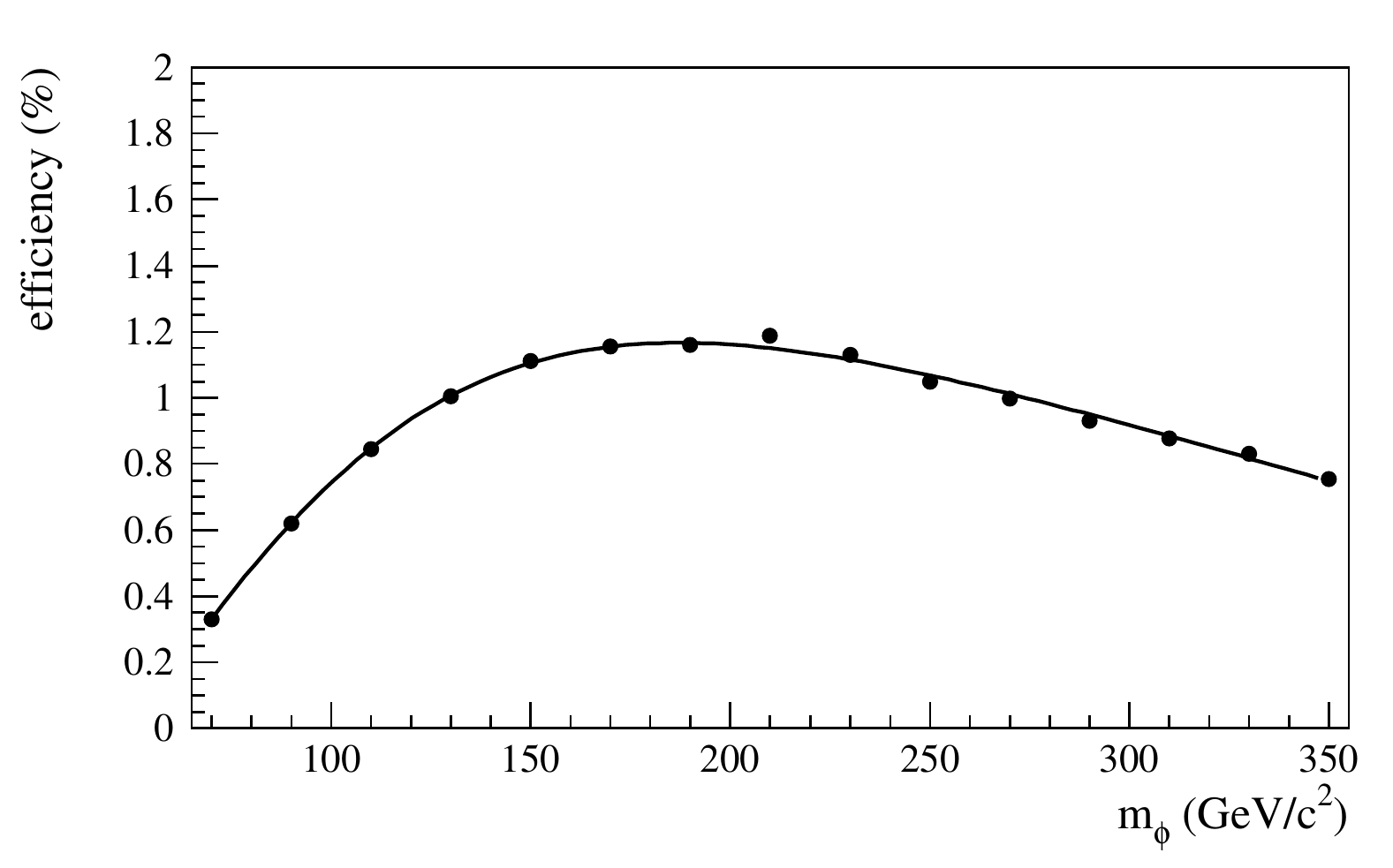}
\end{center}
\end{figure}

\begin{figure}
\caption{Distributions of $m_{12}$ for the simulated signal samples.  
The lines simply connect the bins and do not
represent parametrizations.  All are normalized to unit area.}
\label{f:signal_samples}
\begin{center}
\includegraphics[width=\figwidth]{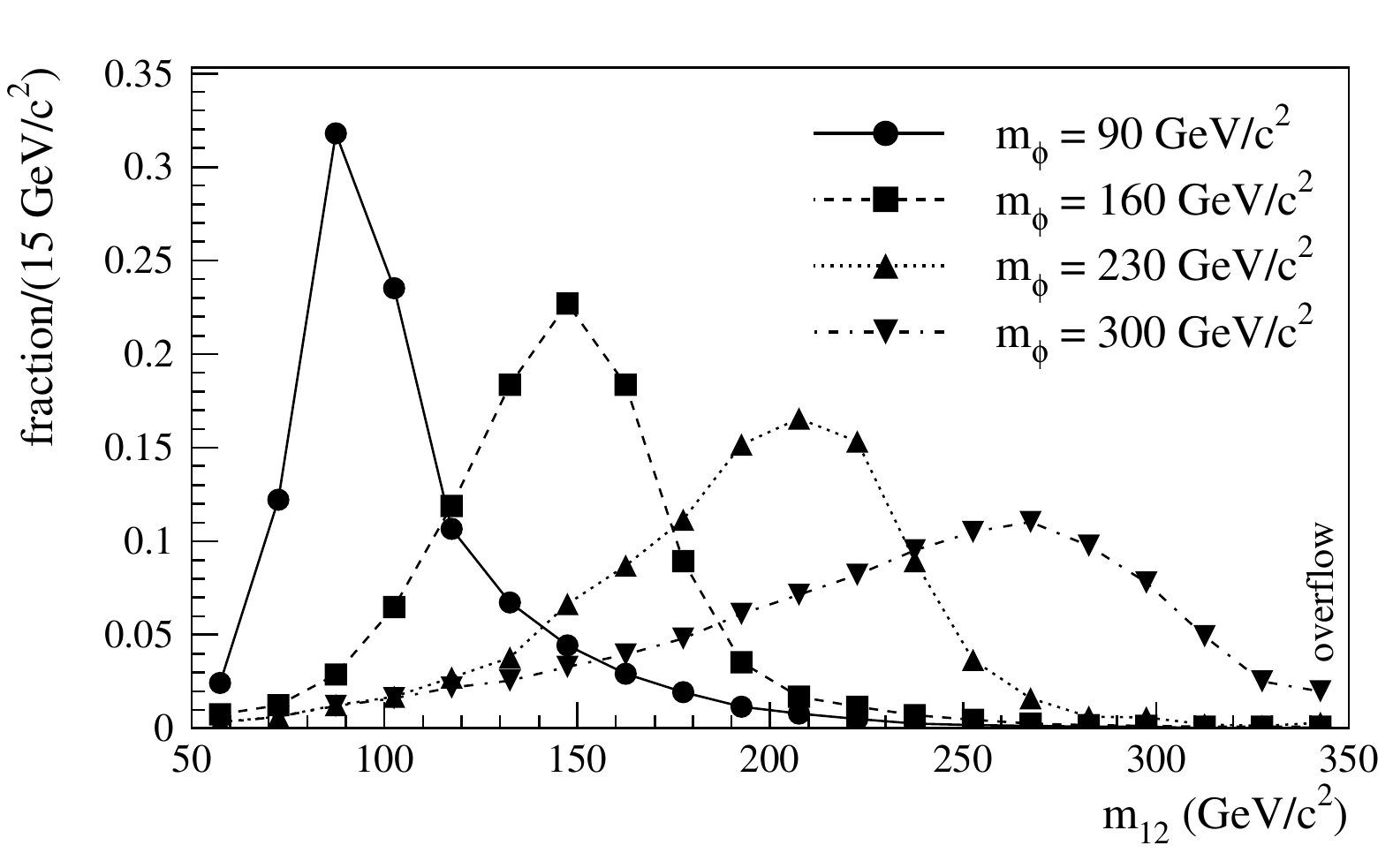}
\end{center}
\end{figure}

The event selection efficiencies vary from 0.3\% to 1.2\% as a
function of the mass of the neutral scalar and are shown in
Fig.~\ref{f:acc_vs_mh}.  The efficiency of the offline requirement
of three or more jets is 14-28\%, the efficiency after adding the
requirement of three or more SECVTX tags is 0.75-1.7\%, and the final
matching requirements of the tagged jets to the trigger clusters and
tracks reduce the efficiency to 0.3-1.2\%.  For a cross section of
10~pb we therefore expect to select 80-310 signal events passing
our requirements.  The mass of the two leading jets in the event
$m_{12}$, which is used to separate signal from background, is shown
in Fig.~\ref{f:signal_samples} for four values of the neutral scalar
mass.

\section{Background Model}
\label{s:bkgd}

Aside from the possible neutral scalar signal, the triple-tagged event
sample is predominantly due to the QCD multijet production of heavy
quarks.  Other processes such as $t\bar{t}$ production and
$Z\rightarrow b\bar{b}$ + jets are found to be negligible, a point to
which we shall return in Section~\ref{s:results}.  The heavy flavor
multijet events arise from a large number of production
mechanisms~\cite{p:lannon} for which the rates are not precisely
known. The differing kinematics of each can produce different $m_{12}$
spectra, which the background estimation must accomodate.  The
$m_{12}$ spectrum of the background is also affected by biases
introduced by the trigger and displaced-vertex tagging requirements.

Heavy quark production can be categorized into three types of
processes~\cite{p:lannon}: flavor creation, flavor excitation, and
gluon splitting.  Flavor creation refers to cases where a pair of
heavy quarks are created directly from the hard scattering process,
i.e. $q\bar{q} \rightarrow b\bar{b} + X$ where the additional activity $X$ in
the event is from initial or final state gluon radiation.  Flavor
excitation refers to processes with a heavy quark in the initial state
which participates in the hard scattering, i.e. $bq \rightarrow bq +
X$.  Cases where the heavy quarks are not directly involved in the
hard scattering are referred to as gluon splitting, i.e. $qg
\rightarrow qg + X$ followed by $g\rightarrow b\bar{b}$ where the 
heavy quark pair is produced as the gluon fragments.  It is possible
to obtain more than two heavy quarks in the final state by combining
these processes in a single event, for example $cg\rightarrow cg + X$
followed by $g\rightarrow b\bar{b}$, or $gg \rightarrow gg + X$ with
both final state gluons splitting into $b\bar{b}$ pairs.  Given the
large number of possible final states with multiple heavy quarks, each
of which can be obtained through a variety of production mechanisms,
estimating the multijet background by direct calculation is a complex
undertaking with potentially large uncertainties.  A
data-driven background estimation of the mixture of processes directly
from the signal sample itself is a more tractable problem, and the
method that we adopt in this analysis.

In order to qualitatively understand which of the many possible heavy
quark final states are necessary to model with our data-driven method,
and to what extent they differ in $m_{12}$, we begin with a study of
simulated samples of generic QCD multijet production.  These samples
are generated using the {\sc pythia} program with MSEL=1
($2\rightarrow 2$ scattering where the outgoing partons can be gluons
or quarks lighter than the top quark) and a simple parametrization of the
secondary vertex tagging efficiency which is a function of the $E_T$,
pseudorapidity, and quark flavor of the jets.  We find in this study
that more than $90\%$ of the QCD background in our selected triple-tag
sample consists of events with at least two $b$-jets, with the
additional tagged jet being any of a mistagged light jet or a
correctly-tagged $c$-jet or third $b$-jet.

In three-jet events with at least two $b$-jets, the additional jet is
also a $b$-jet roughly 2\% of the time, a $c$-jet 4\% of the time,
and a light quark or gluon jet the remaining 94\% of the time.  These
fractions hold when the two $b$-jets are either the two leading jets or
if one of them is the third-leading jet.  The flavor composition of
the additional jet will ultimately be determined by fitting the data
rather than using these estimates, however we will use them as starting
points for the fit and also in the calculation of limits.

We next focus on the $m_{12}$ spectrum in the subset of the {\sc
pythia} generator-level events described above with at least two
$b$-jets.  We compare the spectrum in events with two $b$-jets and at
least one other jet of any flavor to those in events where the
additional jet(s) beyond the inital two $b$-jets has a particular
flavor (charm or another bottom jet).  We find no significant
differences between the flavor-inclusive spectrum and the
flavor-specific ones.  These results hold when splitting the generated
sample by heavy quark production process, so the agreement is general
rather than the result of a cancellation or particular mix of
processes.  Changing the {\sc pythia} hard scattering $Q^2$ scale
factor parameter {\tt PARP(67)} over the range of 1-4 as in
Ref.~\cite{p:lannon} produces significant changes in the $m_{12}$
spectra, however the agreement between the flavor-inclusive and
flavor-specific spectra is preserved as the changes in the underlying
physics affect the two spectra in a similar way.  In order to use {\sc
pythia} directly to estimate the $m_{12}$ spectrum of triple-tagged
events we would need to know the ``correct'' values of {\tt PARP(67)}
and other parameters, but the similarity in $m_{12}$ shape between
double-tagged (flavor-inclusive) and triple-tagged (flavor-specific)
events appears to be insensitive to the details of any particular {\sc
pythia} tuning.  All of the relevant jet physics is therefore already
contained in the double-tagged sample, which can be selected from data
to remove dependence on event generators such as {\sc pythia}.  The
only correction necessary to use the double-tagged events as
background estimates for the triple-tagged sample is the purely
instrumental bias of requiring the third tag.

Based on the results of the generator-level study, we conclude that
$b\bar{b}$ plus a third tagged jet of any flavor represents more than $90$\%
of the heavy flavor multijet background.  This is the basis of our
background model; the effect of neglecting the $10$\% component with fewer
than two $b$-jets is discussed in Sec.~\ref{s:results}.
Because the properties of the additional jets in $b\bar{b}$ events do
not depend strongly on the flavor of the jets, we can use the sample
of double-tagged events described in Sec.~\ref{s:data} as a
representation for all possible flavors of the third tags.  

The efficiency of requiring the third tag does depend upon the flavor,
so we construct background estimates which depend on the flavor of the
jet and its position in the $E_T$-ordered list of jets in the event.
Splitting the background estimates in this way also provides
flexibility to accomodate mixtures of production processes.  For
example, events where the two leading jets are both $b$-jets are more
likely to result from flavor creation of $b\bar{b}$ than are events
with the second- and third-leading jets both $b$-jets, which have a
larger contribution from a gluon splitting to $b\bar{b}$ and recoiling
against another parton from the hard scatter.  The normalizations of
these flavor- and topology-dependent estimates will be determined from
a fit to the data so as to minimize dependence on theoretical inputs.

In the remainder of this section we show how we estimate the heavy
quark multijet background from the large sample of data events with two
$b$-tags. 

\subsection{The Double-Tagged Sample}

That the triple-tagged sample predominantly contains at least two
$b$-jets is of major importance. First, it reduces the number of
flavor combinations which must be considered to determine its
composition to a manageable level. Secondly, samples of $b\bar{b}$
events with at least one additional jet are easily selected from the
same dataset as the signal region and are therefore subject to the
same biases from the trigger and displaced-vertex tagging of the two
$b$-jets as the events in the signal region. By simulating the effect
of the SECVTX tag on the third jet, we can use the double-tagged
sample to model all components of the triple-tagged sample with two or
more $b$-jets.  Because we are going to determine the normalizations
from a fit to the data, we need only to model the shape of the
$m_{12}$ spectrum for each background component.

\begin{figure}
\caption{Efficiency to tag a jet with the indicated flavor as a
function of the jet $E_T$ (a) and the number of tracks passing SECVTX
quality requirements in the jet (b).  Only jets with at least two
quality tracks are included.  The highest bin in each plot includes
overflows.}
\label{f:tagrates}
\begin{center}
\includegraphics[width=\figwidth]{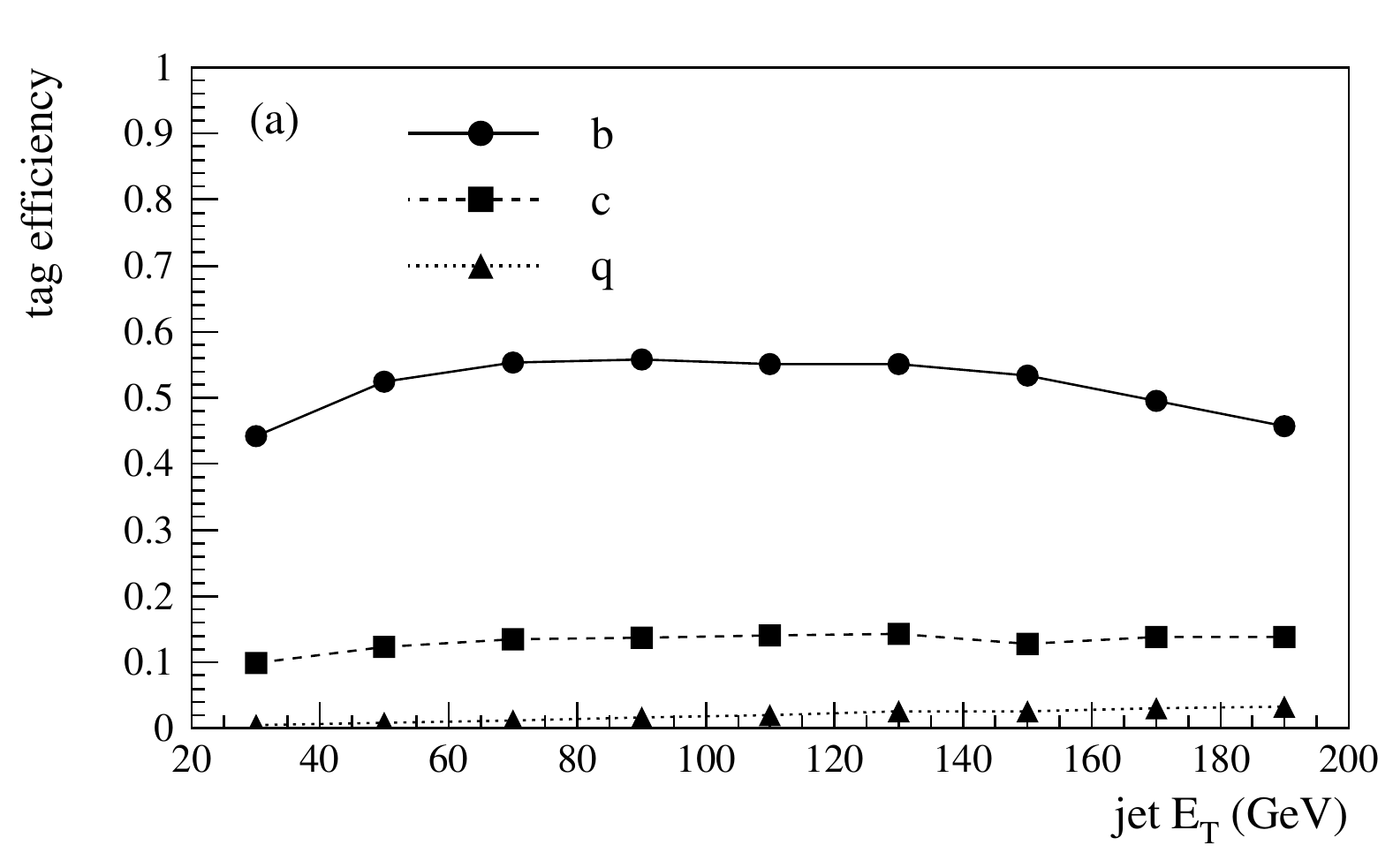}
\includegraphics[width=\figwidth]{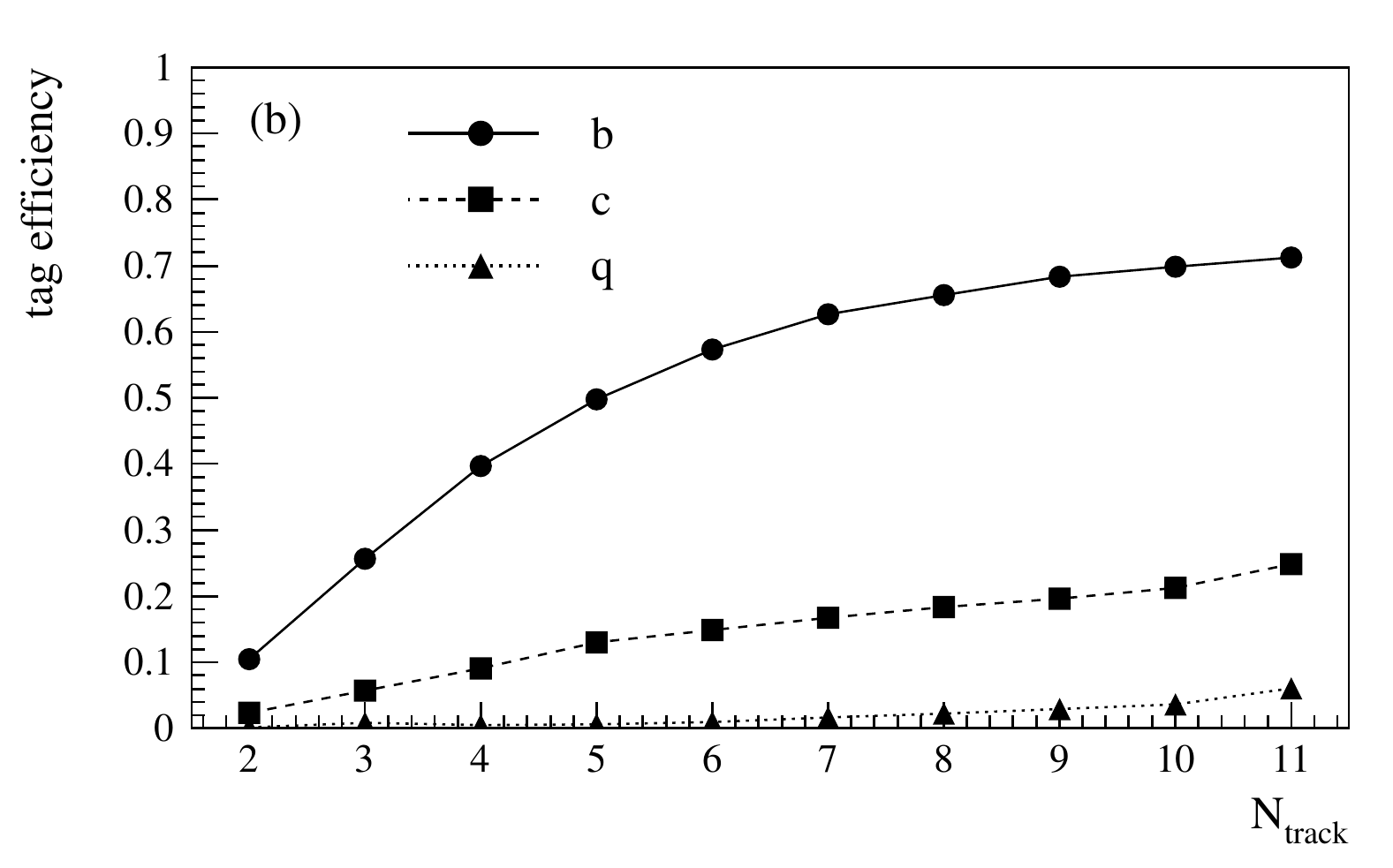}
\end{center}
\end{figure}

For moderate values of jet $E_T$ SECVTX becomes more efficient as jet
$E_T$ increases, particularly for light-flavor jets where the false
tag rate is highly dependent upon the number of candidate tracks in
the jet which scales as the jet $E_T$.  For $b$ and $c$ quark jets the
effect is less dramatic, and does not hold over the full range of
$E_T$.  This effect is illustrated in Fig.~\ref{f:tagrates}.  The drop
in efficiency for $b$ quark jets at higher $E_T$ is due to increasing
track occupancy in the jets, which causes the silicon tracker to merge
hits from different tracks resulting in lower-quality tracks which are
rejected by the SECVTX requirements.  Because of these variations of
the efficiencies as a function of jet $E_T$, requiring SECVTX tagged
jets will bias the events to a different $m_{12}$ spectrum than is
observed in the parent, untagged sample.  The double-tagged sample
which is the starting point for our background estimates already
includes the bias due to the two existing tags, so we must simulate
only the bias which would be due to requiring the third tag as in the
signal region.  This is accomplished by weighting the events using
efficiency parametrizations for $b$, $c$, and light-flavor jets
derived from large samples of fully-simulated {\sc pythia} multijet
events.  The efficiencies are parametrized as a function of the jet
$E_T$ and the number of tracks in the jet passing the SECVTX quality
cuts.  As these efficiencies are derived from simulated samples, they
are corrected to match the $E_T$-parametrized efficiencies observed in
the data using the same procedure employed for the simulated Higgs
boson samples.

We describe the flavor structure of the jets in the event in the form
$xyz$, where $xy$ denotes the flavor of the two leading jets and $z$ is the
flavor of the third-leading jet or fourth-leading jet in the case that
the third-leading jet is not tagged by SECVTX.  For example, $bqb$
would denote events where the two leading jets are a $b$-jet ($b$) and
a mistagged light quark (or gluon) jet ($q$), and the third tag is
another $b$-jet.  Because our search variable $m_{12}$ is symmetric
under the interchange of the two leading jets, we make no distinction
between the leading and second-leading jets so that in a $bqb$ event
the gluon or light-flavor jet $q$ could be either of the two leading
jets.

With this convention, we identify five types of event with at least
two $b$-jets.  Three involve $b$-jets in both of the leading jets:
$bbb$, $bbc$, and $bbq$.  The other two, $bcb$ and $bqb$, have the
non-$b$-jet in one of the two leading jets.  The distinction between
the flavor content within the two leading jets and the flavor of the
third jet is important, as the events will have differing kinematics
and tagging biases when comparing $bbq$ {\it vs.} $bqb$.  In $bbq$
events the two $b$-jets are likely to have originated directly from
the hard scatter, while in $bqb$ it is more likely that the two
$b$-jets come from a gluon splitting as mentioned above.  The SECVTX
algorithm is much more biased toward high-$E_T$ jets for light flavor
than it is for $b$-jets, so we expect that $bqb$ events will have a
harder $m_{12}$ spectrum than $bbq$.  Because we do not want to make
any assumption about the rate of gluon splitting relative to
$b\bar{b}$ flavor creation, we use both estimates and allow the fit of
the data to determine the relative proportions.

\subsubsection{Corrections to the Double-Tagged Sample}

While our model assumes two $b$-jets in each event, the
generator-level study described above indicates that the double-tagged
events have a contribution of $\sim 10$\% where one or both of the
tagged jets is a ``mis-tagged'' light flavor jet.  We correct for this
using events which have two displaced vertices, but where one or both
of the vertices are on the opposite side of the primary vertex from
the jet direction.  These ``negative'' tags are predominantly fake
tags from light-flavor jets and are a product of the finite position
resolution of the tracking system.  We expect there to be an equal
number of fake tags from this source on the default, ``positive''
side, together with additional contributions of fake tags from
$K_S/\Lambda$ and interactions with the detector material which are
not present in the negative tags.  The negative tags also contain a
small contribution from heavy-flavor jets which should be subtracted
in order to obtain the positive fake rate.  The total number of
positive fake tags is found by scaling the negative tag count by a
factor $\lambda = 1.4 \pm 0.2$~\cite{p:secvtx} which accounts for the
effects described above and is measured from the data.  We find no
significant variation of $\lambda$ as a function of jet $E_T$.

We weight these events to simulate the third tag in the same way as
the events with two default ``positive'' tags and then compute the
number of true $b\bar{b}$ events using

\begin{equation}
\label{e:nonbb}
N_{b\bar{b}} = N_{++} - \lambda N_{+-} + \lambda^2 N_{--}
\end{equation}

\noindent where $N_{++}$ is the number of observed positive
double-tags, $N_{+-}$ is the number of events with one of the tags
negative, and $N_{--}$ is the number with both tags negative.  This
relation can be understood by considering $N_{+-}$ as the number of
events with either one $b$-tag and one fake tag or two fake tags.  The
two fake tag case will be double-counted by this estimate, because
there are two permutations for which jet is the positive tag and which
is the negative tag.  Therefore the $N_{--}$ term which is an estimate
of the number of two fake tag events is added to correct for the
double-counting.  The $\lambda$ factors are inserted to correct the
negative tag rates into estimates of the total positive fake tag
rates.

\begin{figure}
\caption{Distributions of $m_{12}$ used to construct the corrected $bcb$ 
background estimate.  The +c+ shape (the initial estimate with two
positive tags, before correction) is shown with unit area.  The
+c-/-c+ shapes (starting from one positive and one negative tag) and
-c- shapes (two negative tags) are shown with normalizations
proportional to their area compared with +c+.  For -c- a further
scaling by a factor ten is applied to enhance visibility.  The
corrected estimate is reduced in area by $\sim$10\% relative to +c+.}
\label{f:nonbb-corr-bcb}
\begin{center}
\includegraphics[width=\figwidth]{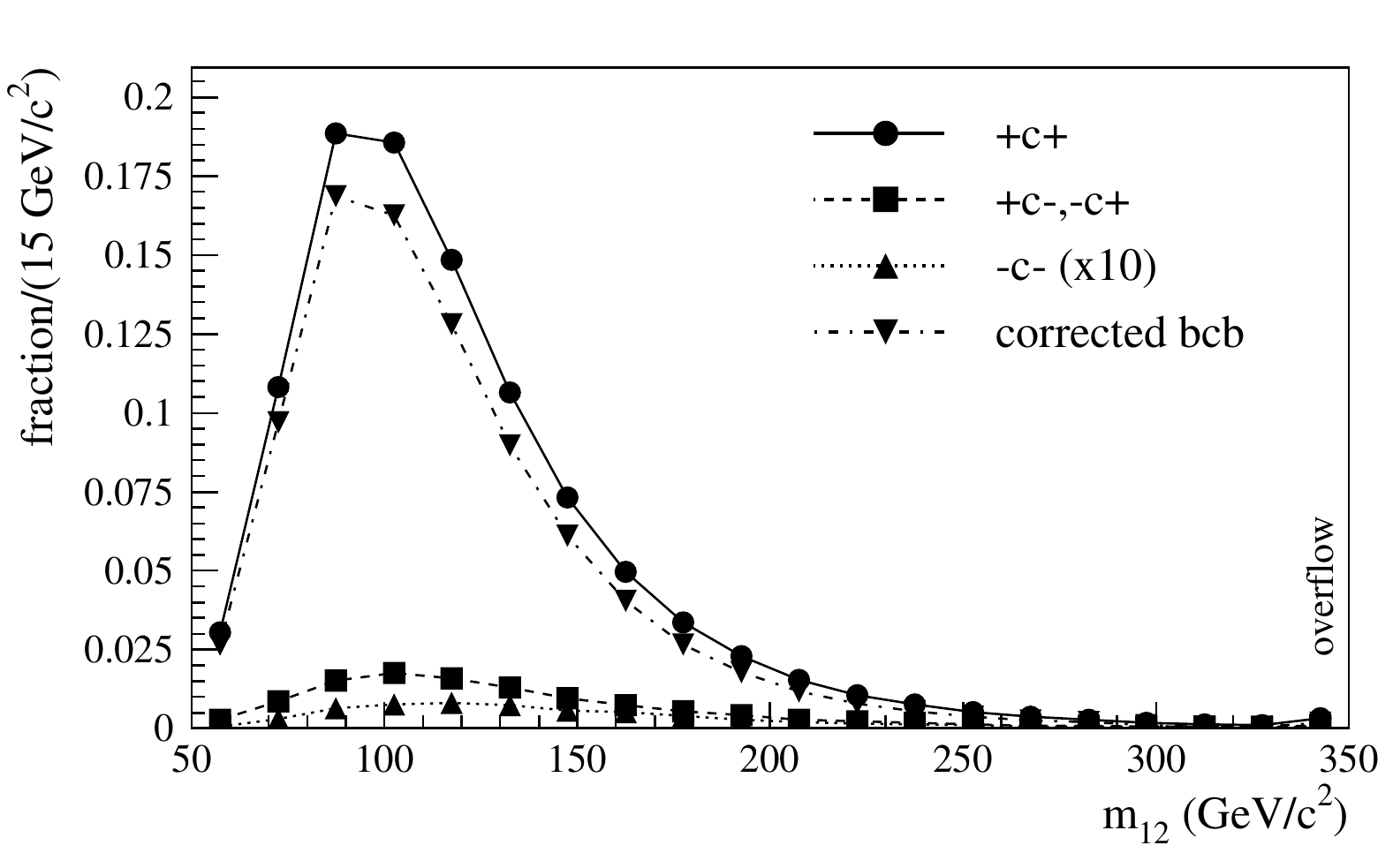}
\end{center}
\end{figure}

This correction to subtract the non-$b\bar{b}$ component is applied
bin-by-bin in $m_{12}$ when constructing estimates for all five of
the background components.  It reduces the normalization by around
10\% and also softens the $m_{12}$ spectrum in each estimate, because
the samples with one or two negative tags will have harder $m_{12}$
spectra than the sample with two positive tags due to the fake tag
bias towards higher jet $E_T$ effect described above.  The effect of
the correction is illustrated in Fig.~\ref{f:nonbb-corr-bcb} for the
$bcb$ background estimate.

\subsection{The Heavy Flavor Multijet Background Components}

We now describe in detail how each of the five model components, or
``templates'', for the three-tag backgrounds is constructed from the
double-tag data. When referring to the templates we adopt the
convention of capitalizing the assumed flavor of the untagged jet, so
that for the $bbq$ background we would denote the template as $bbQ$.
This distinction is most important for the $bbb$ background as will be
seen later.

\boldmath
\subsubsection{The $bbc$ and $bbq$ backgrounds}
\unboldmath

Starting from the corrected double-tagged sample with the two leading
jets tagged, we weight the events by the probability to tag the third
jet if it were a $c$-jet or a light-quark jet to produce estimates for
the $bbc$ and $bbq$ background components, respectively. If a fourth
jet exists we add the weights to tag either the third or fourth jet.

\boldmath
\subsubsection{The $bcb$ and $bqb$ backgrounds}
\unboldmath

The templates for these backgrounds are constructed in essentially the
same way as $bbC$ and $bbQ$.  The difference is that we start from a
double-tagged sample where one of the tags is in the third or fourth
jet rather than requiring that both of the two leading jets be tagged
as in $bbQ/bbC$.  From there we subtract the non-$b\bar{b}$ component
using Eq.~\ref{e:nonbb} and weight the untagged jet within the two
leading jets with either the charm-tag efficiency or the light-flavor
jet mistag probability.  The event selection requires that there be at
least two level~2 trigger silicon tracks matched to the two leading
jets, so for example in the $bbQ$ template we require that the two
leading jets contain at least two matched level~2 tracks (either at
least one in each jet or at least two in one of the jets).  For the
$bCb$ and $bQb$ templates we require that only one of the two leading
jets be tagged, and simulate the tag in the other jet.  If the tagged
jet has fewer than two matched level~2 silicon tracks, we use an
efficiency parametrization for the other of the two leading jets that
represents not only the efficiency to tag the jet with SECVTX (as is
used in the $bbQ$ case, for example, to simulate the fake light-flavor
tag of the third or fourth jet) but also for that jet to contain
enough matched level~2 silicon tracks so that the total for the two
leading jets is at least two.  So, for example, if the leading jet is
tagged and has one matched level~2 silicon track, we would weight the
event by the combined efficiency to not only tag the second-leading
jet with SECVTX but also to have matched at least one level~2 silicon
track to it.  In this way the effect of requiring at least two matched
level~2 silicon tracks within the two leading jets is modeled.  In the
example with one matched track in the leading jet, there must be a
second level~2 silicon track somewhere in the event for it to have
passed the online trigger selection.  We account for this by requiring
that between the two tagged jets (the leading jet and either the
third- or fourth-leading in our example) there must be at least two
matched level~2 silicon tracks.  The requirement of the matched track
in the third- or fourth-leading jet is not present in the signal
sample, so this represents an unwanted bias.  We remove the bias by
additionally weighting these events by the ratio of the inclusive
SECVTX $b$-jet tag efficiency for the third or fourth jet to the
efficiency for SECVTX tagging with matched level~2 tracks.

\boldmath
\subsubsection{The $bbb$ background}
\unboldmath

The third-tag weighting procedure works straightforwardly for the $bbc$
and $bbq$ backgrounds, because the $b$-quark production physics is the
same as in the $bbj$ events used as the starting point: the $b$-jets
in the double-tagged sample can be mapped directly to the signal
region and the various $b\bar{b}$ production mechanisms are properly
represented.  For the $bbb$ background this is not the case, because
there are two $b\bar{b}$ pairs present.  Sometimes the two leading
jets in the event are from the same $b\bar{b}$ pair, in which case a
$bbB$ template would be the appropriate choice because it is derived
from events with the $b\bar{b}$ pair in the two leading jets.  Other
times the two leading jets are from a different $b\bar{b}$ pair, where
a $bBb$ template would be a better representation.

The two methods of constructing a template for $bbb$ have
significantly different $m_{12}$ distributions, which is due to the
particular kinematics of $b\bar{b}$ production through gluon
splitting.  Gluon splitting produces $b\bar{b}$ pairs which tend to be
less back-to-back than other production mechanisms.  When the two
$b$-jets in such an event are the two leading jets, as in the $bbB$
template, we observe a softer $m_{12}$ distribution than is seen in
$bBb$, where only one jet from the $b\bar{b}$ pair is within the two
leading jets and the other of the two leading jets is an additional
jet in the event against which the $b\bar{b}$ system is recoiling.

\begin{figure}
\caption{Distributions of $m_{12}$ from the generator-level {\sc
pythia} study, for simulations of the double-tagged background
templates $bBb$ and $bbB$, compared to events with three true
$b$-jets.}
\label{f:bbb_bracket}
\begin{center}
\includegraphics[width=\figwidth]{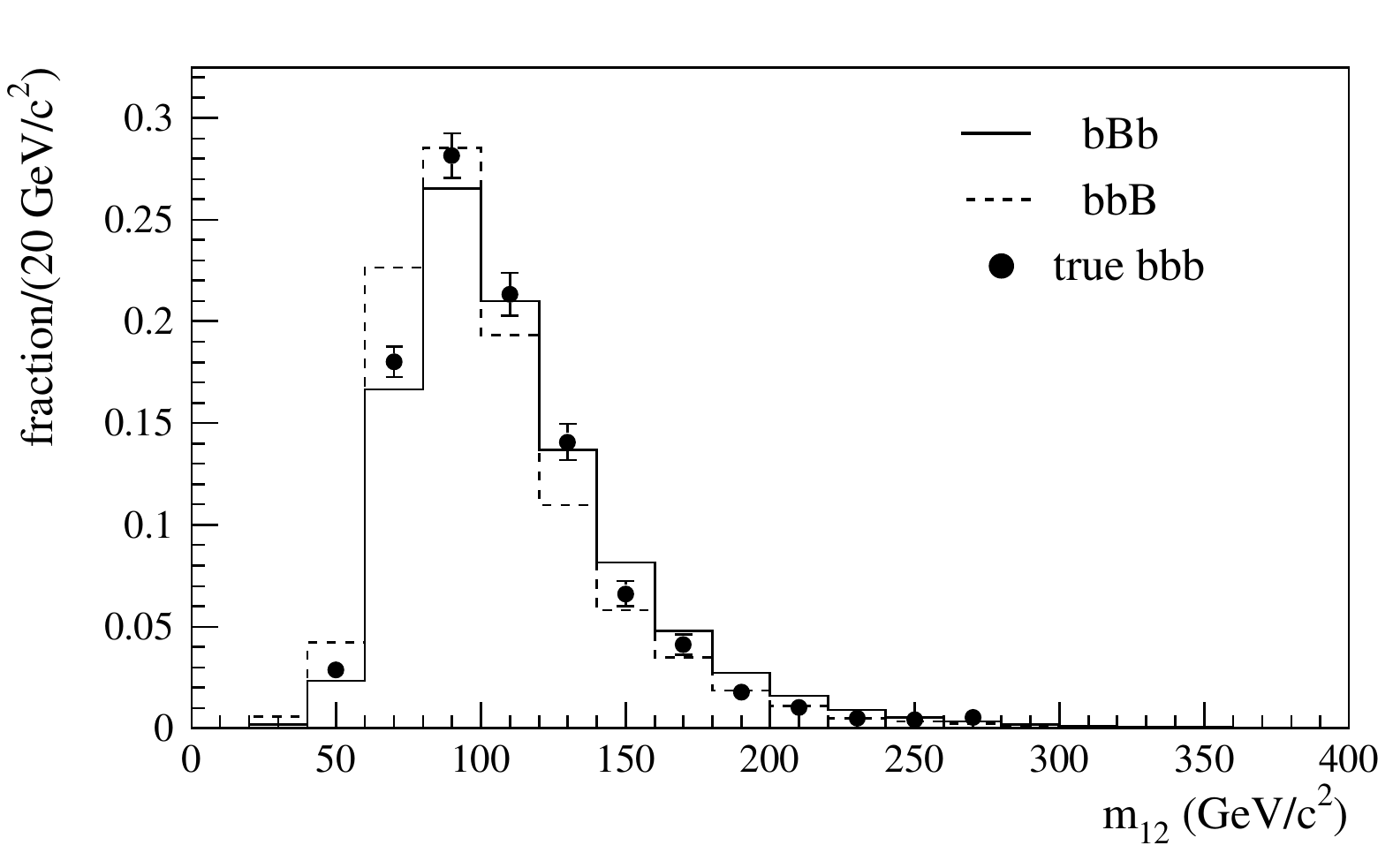}
\end{center}
\end{figure}

The {\sc pythia} simulation indicates that the $m_{12}$ spectrum for
$bbb$ events lies between the two estimates $bbB$ and $bBb$, as shown
in Fig.~\ref{f:bbb_bracket}.  The
difference between the two estimates is largest for events involving
only gluon splitting, but the relationship also persists across other
heavy-flavor production mechanisms.  We conclude that regardless of
the relative rates of $b\bar{b}$ production processes, the $bbb$
background can be derived from an interpolation between the two
templates $bbB$ and $bBb$.  We include both in the fit and
let the data determine the proper weighting.

\subsubsection{Backgrounds summary}

\begin{figure}
\caption{Distributions of $m_{12}$ for the background fit templates.
The lines simply connect the bin centers and do not represent
parametrizations.  The $bBb$ template is obscured by $bCb$ because
they have nearly the same $m_{12}$ distribution.  All are normalized
to unit area.}
\label{f:bkgd_shapes_data}
\begin{center}
\includegraphics[width=\figwidth]{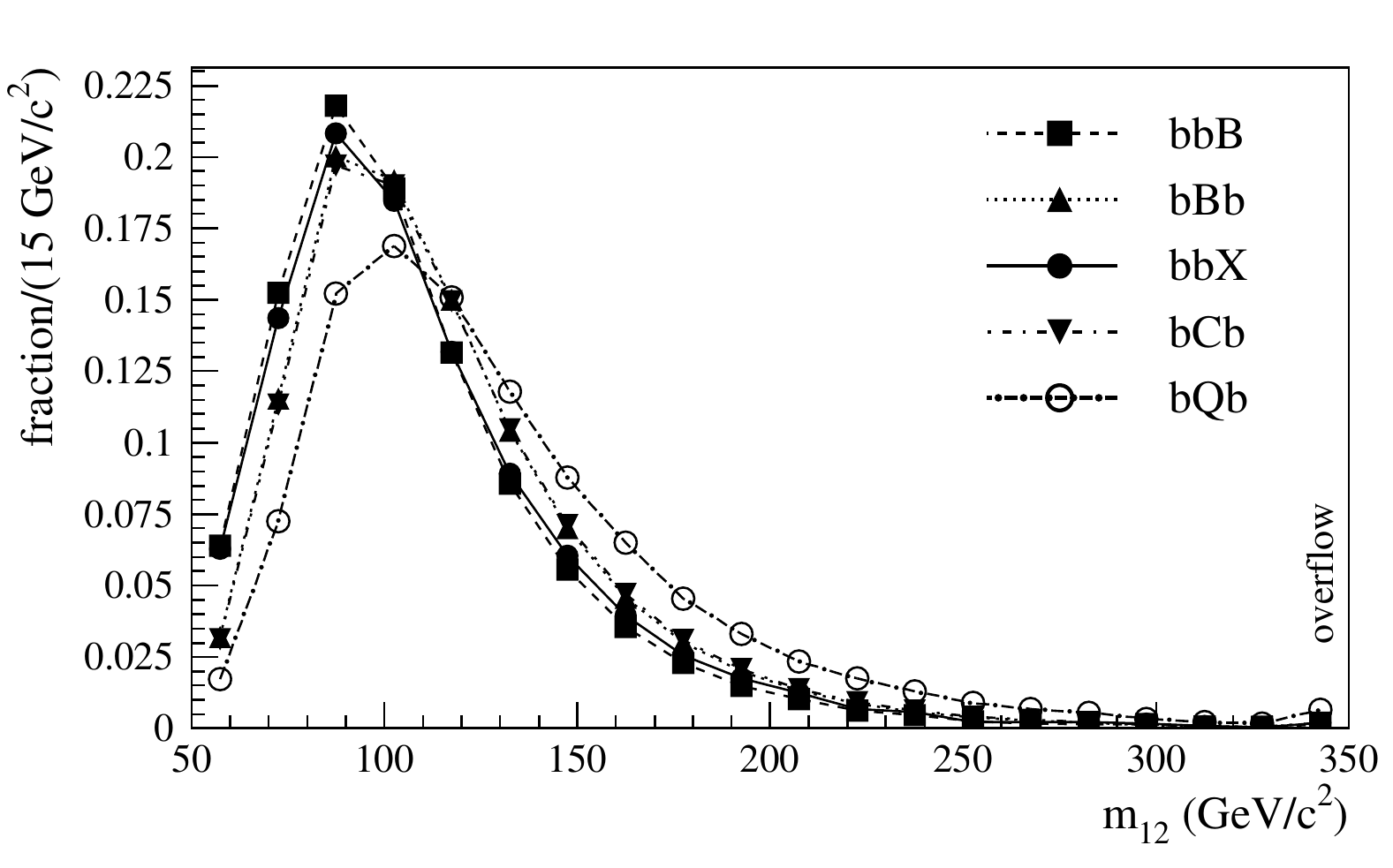}
\end{center}
\end{figure}

The full set of background fit templates for $m_{12}$ is shown in
Fig.~\ref{f:bkgd_shapes_data}.  Because they are too similar to
discriminate in the fit, we use an average of the $bbC$ and $bbQ$
templates which we denote $bbX$.  The backgrounds with two heavy
flavor jets in the leading jet pair have similar $m_{12}$
distributions.  Because the false tag rate rises with jet $E_T$ more
rapidly than does the $b$-tag or $c$-tag rate, the $bQb$ displays a
harder spectrum than $bCb$ or $bBb$ even though they are derived from
the same events. 

\subsection{Fitting the Model to the Data}

Our search will examine the $m_{12}$ distribution for an
enhancement riding atop the continuum background. The search will be
done using a simultaneous fit for the normalization of six
distributions: one neutral scalar model of varying mass, and the five
background templates that together will model the background.  A fit
in $m_{12}$ alone is challenged by the fact that the background
templates peak at similar mass but have different widths, as seen in
Fig.~\ref{f:bkgd_shapes_data}.  A possible signal riding on the
falling edge of the background above 100~GeV/$c^2$ could therefore be
fitted by adding additional contribution from the wide $bQb$
distribution, for example, resulting in loss of sensitivity.  This
effect can be mitigated by adding another variable which is sensitive
to the differing flavor content of the templates, which we call
$x_{tags}$. In this section we describe the $x_{tags}$ variable and
then examine the ability of our background model alone to describe the
data without any contribution from the neutral scalar signal model, using a
two-dimensional fit of the distributions of $m_{12}$ and $x_{tags}$.

\subsubsection{The Flavor Dependent Variable $x_{tags}$}

Because we are going to fit the $m_{12}$ spectrum of the triple-tagged
data with our background templates, each of which has its own
characteristic $m_{12}$ spectrum, it is useful to have a second method
with which to constrain the relative fractions of each background
template and obtain a firmer prediction of the overall background
$m_{12}$ spectrum. The $x_{tags}$ variable should be sensitive to the
flavor of the tagged jets using information independent of $m_{12}$.

\begin{figure}[tp]
\caption{The tag mass $m_{tag}$ for different jet flavors, from the CDF 
simulation.  All distributions are normalized to unit area.}
\label{f:tag_mass}
\begin{center}
\includegraphics[width=\figwidth]{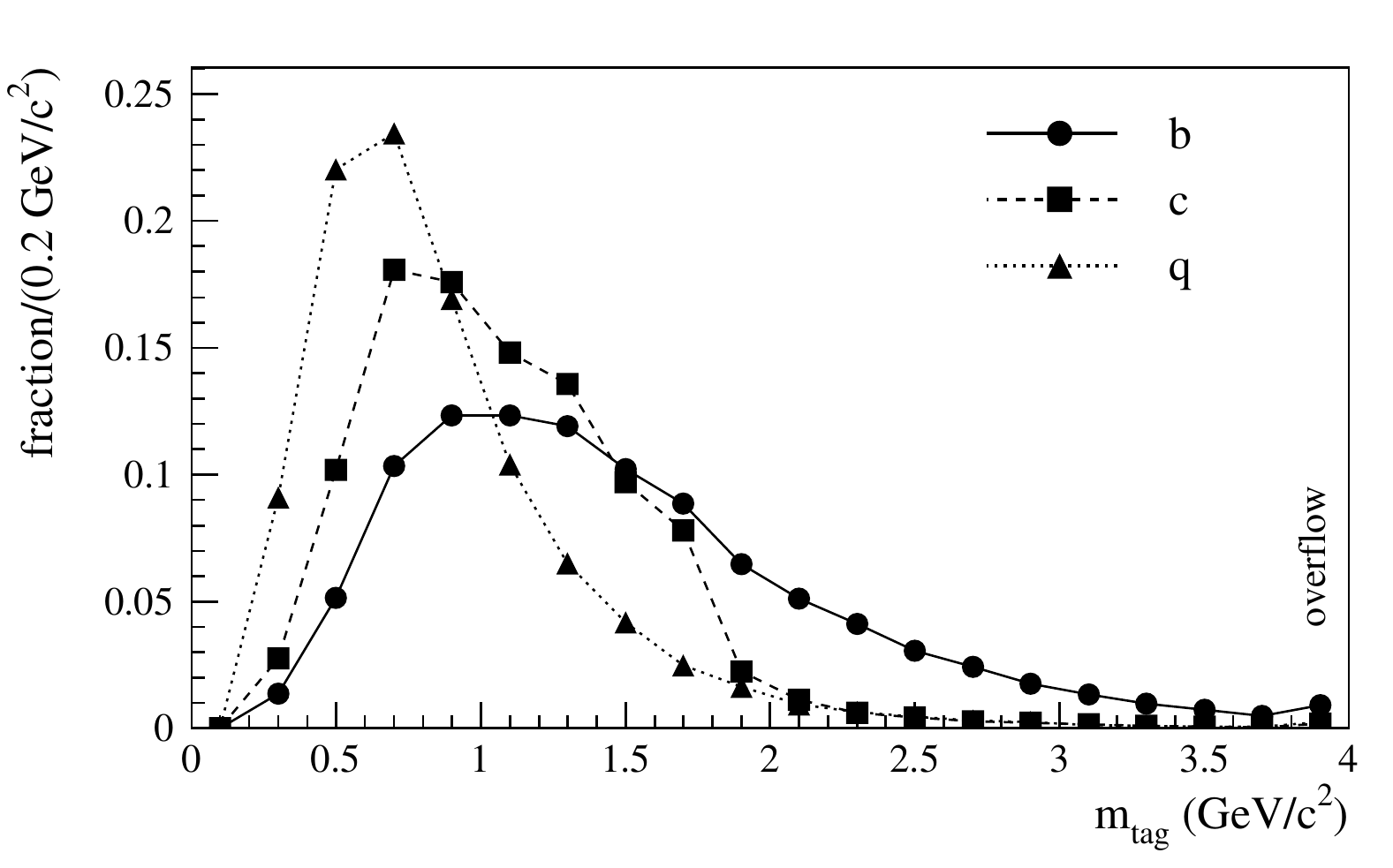}
\end{center}
\end{figure}

The observable chosen as the basis of $x_{tags}$ is $m_{tag}$, the
invariant mass of the tracks which constitute the secondary vertex as
determined by SECVTX. This reflects the masses of the underlying heavy
flavor hadrons and is sensitive to the flavor of the jet as shown in
Fig.~\ref{f:tag_mass}.  We define the quantity
$x_{tags}(m_{1,tag},m_{2,tag},m_{3,tag})$, where $m_{i,tag}$ is the
mass of the tracks forming the displaced vertex in jet 1, 2, or 3.

For example, as mentioned above we expect that $bqb$ events will
exhibit a harder spectrum than $bbq$ due to the bias from the fake tag
in one of the two leading jets.  The $x_{tags}$ variable
should therefore be constructed so that it is sensitive to the
presence of a charm or fake tag in one of the two leading jets, using
$m_{1,tag}$ and $m_{2,tag}$.  If these events were removed, we would
be left with backgrounds where the two leading jets are both $b$-jets
and the third leading jet is any flavor.  The case where the third jet
is also a $b$-jet constitutes an irreducible background to the
potential neutral scalar signal in the $x_{tags}$ spectrum, because the
signal is also three $b$-jets.  However, the backgrounds where the
third jet is a charm or fake tag ($bbc$ and $bbq$) can be separated
from the $bbb$ cases using $m_{3,tag}$.

\begin{figure}[tp]
\caption{Illustration of the $x_{tags}$ definition.  All axes are in units 
of GeV/$c^2$.}
\label{f:xtags-def}
\begin{center}
\includegraphics[width=\figwidth]{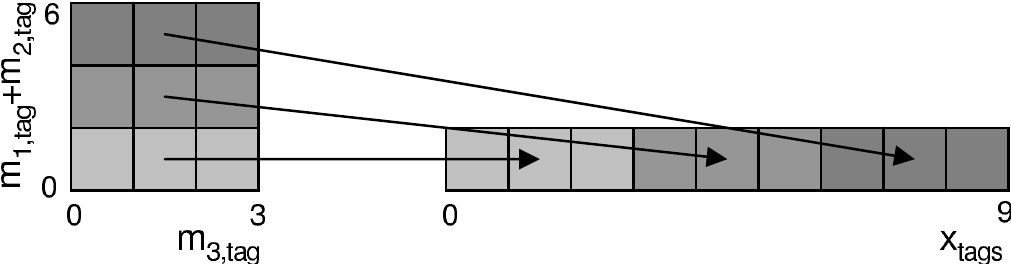}
\end{center}
\end{figure}

Because we make no distinction between the two leading jets in our
flavor classification scheme, we construct $x_{tags}$ to be symmetric
under their interchange, as is $m_{12}$.  We are interested only in
the flavor combination of the pair.  We choose a simple sum
$m_{1,tag}+m_{2,tag}$ to satisfy this constraint.  Combined with the
information from $m_{3,tag}$ we have a two-dimensional distribution,
however we want to reduce this to a single variable so that when
combined with $m_{12}$ we are left with two-dimensional fit templates.
To this end we define the $x_{tags}$ variable as

\begin{equation}
x_{tags} = \left\{ \begin{array}{c@{\quad:\quad}l} \min(m_{3,tag},3) &
m_{1,tag}+m_{2,tag} < 2 \\ \min(m_{3,tag},3)+3 & 2 \leq
m_{1,tag}+m_{2,tag} < 4 \\ \min(m_{3,tag},3)+6 & m_{1,tag}+m_{2,tag}
\geq 4 \end{array} \right\}
\end{equation}

\noindent where $\min(a,b)$ returns the minimum of $a$ and $b$, and
all quantities are in units of GeV/$c^2$.  The net effect is to
unstack a two-dimensional histogram of $m_{1,tag}+m_{2,tag}$ versus
$m_{3,tag}$ into the one-dimensional variable $x_{tags}$, as
illustrated in Fig.~\ref{f:xtags-def}.  The $m_{1,tag}+m_{2,tag}$
axis provides the sensitivity to $bcb$ and $bqb$ versus the other
backgrounds, and the $m_{3,tag}$ separates out $bbc$ and $bbq$.

\begin{figure}
\caption{Projections onto the $m_{tag}$ axis of the tag probability
parametrization, for jets with $80 < E_T < 100$~GeV and the indicated
numbers of quality SECVTX tracks, for $b$ jets (a) and
fake tags of light-flavor jets (b).  The area of each
histogram indicates the total tag probability for this slice of $E_T$
and number of tracks.}
\label{f:mtag_scan}
\begin{center}
\includegraphics[width=\figwidth]{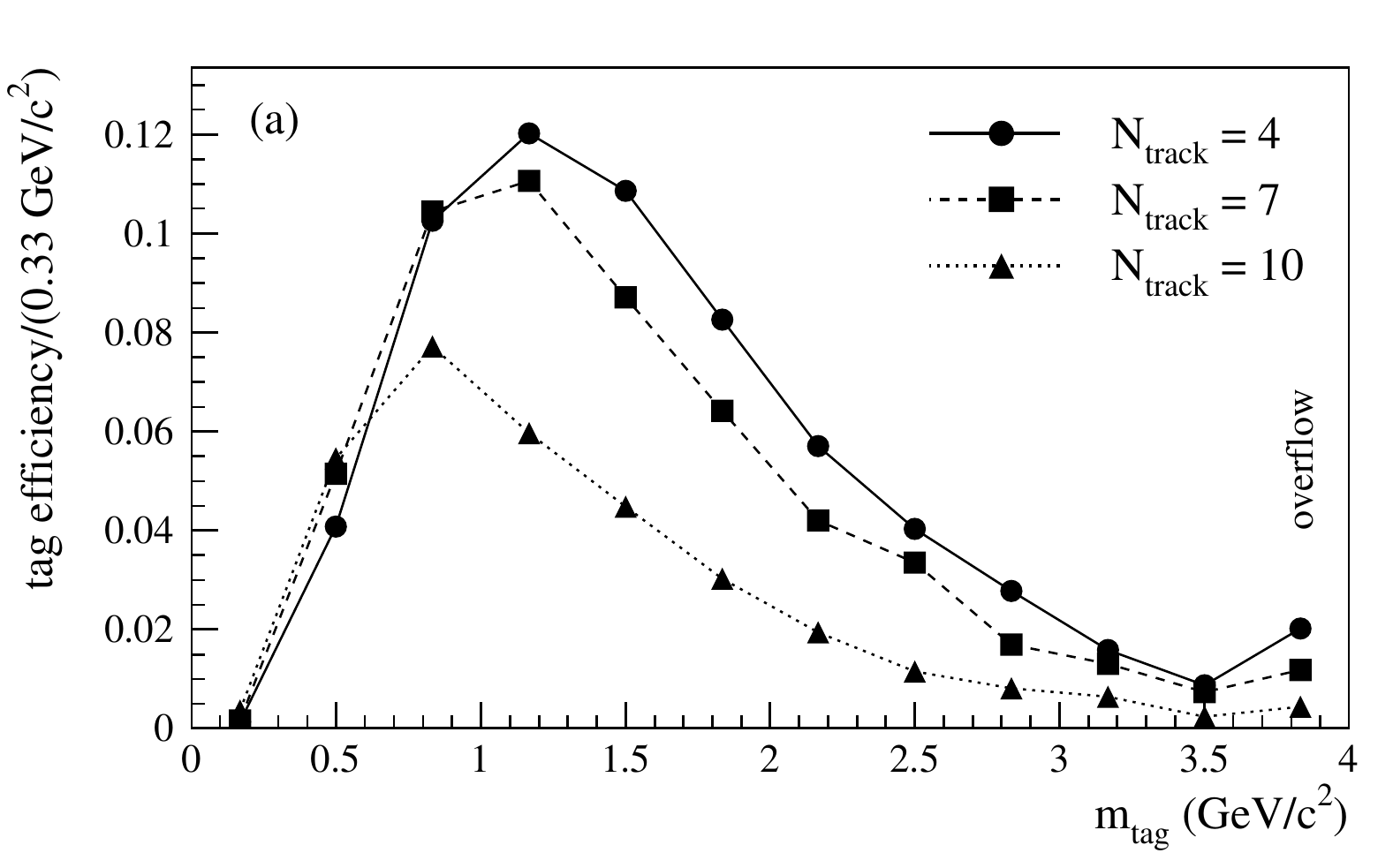}
\includegraphics[width=\figwidth]{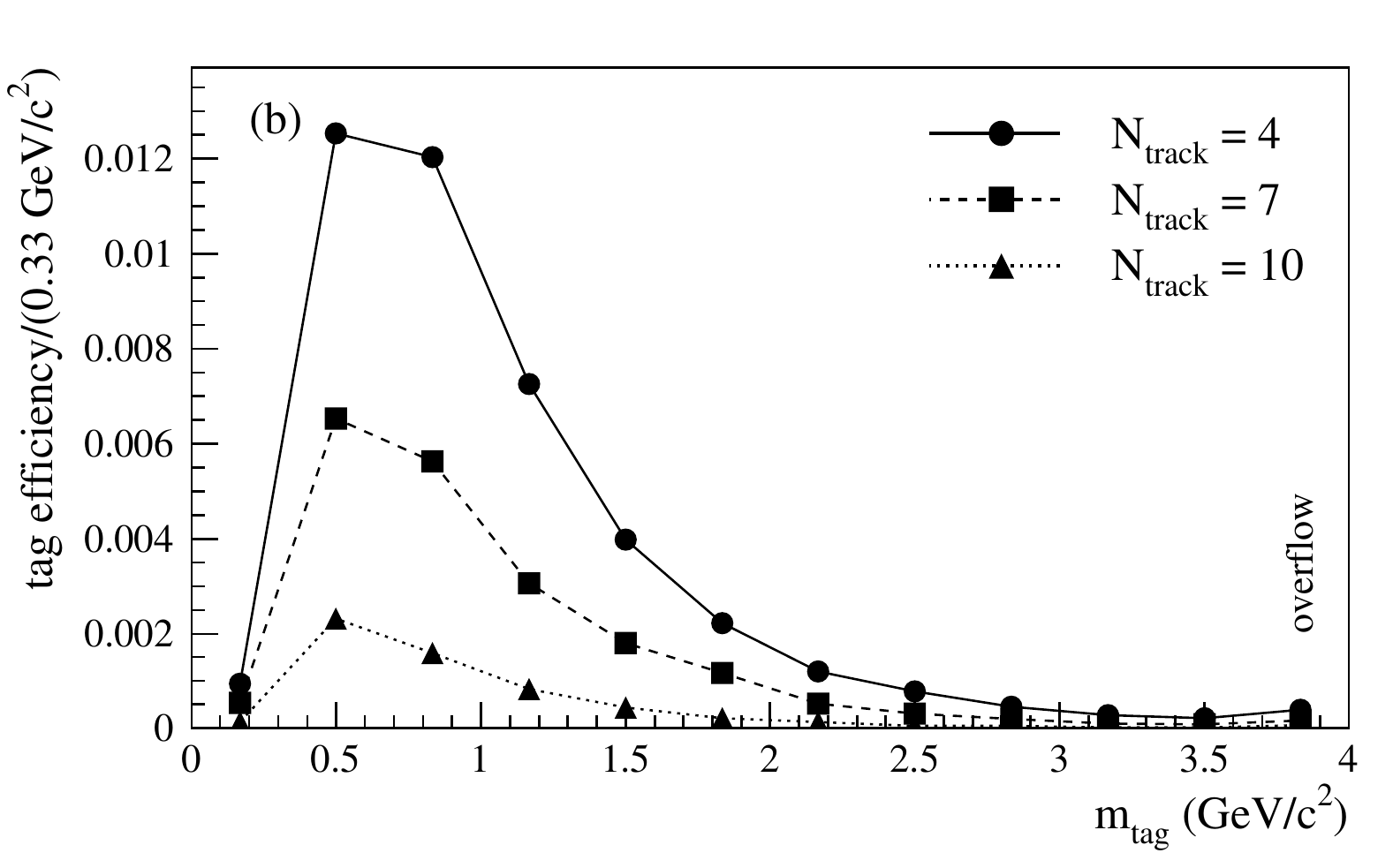}
\end{center}
\end{figure}

In order to compute $x_{tags}$ for the background templates we need to
simulate not only the efficiency of the third tag for each event, but
also its expected $m_{tag}$ spectrum.  This is done by extending the
tag efficiency parametrization so that it is a function of the jet
$E_T$, the number of quality tracks, and $m_{tag}$.  The
parametrization can then be considered to represent the probability to
tag a jet with a given $E_T$ and number of quality tracks and an
assumed flavor of $q$, $c$, or $b$, and for that tag to have a
particular tag mass $m_{tag}$.  Projections of this parametrization
onto the $m_{tag}$ axis for particular values of $E_T$ and number of
tracks are shown in Fig.~\ref{f:mtag_scan}.  For each event we
iterate over all bins of $m_{tag}$ in the parametrization for the
simulated third tag, compute the corresponding $x_{tags}$ for each
bin, and build up the background templates using the parametrization
to estimate the approriate weight for each value of $m_{tags}$ in the
third tag.  Each event will contribute to only a single bin in
$m_{12}$ but can fill multiple bins in $x_{tags}$ as we iterate over
the bins of $m_{tag}$ for the third tag.

\begin{figure}
\caption{Distributions of $x_{tags}$ for the background fit
templates.  All are normalized to unit area.}
\label{f:bkgd_shapes_xtags}
\begin{center}
\includegraphics[width=\figwidth]{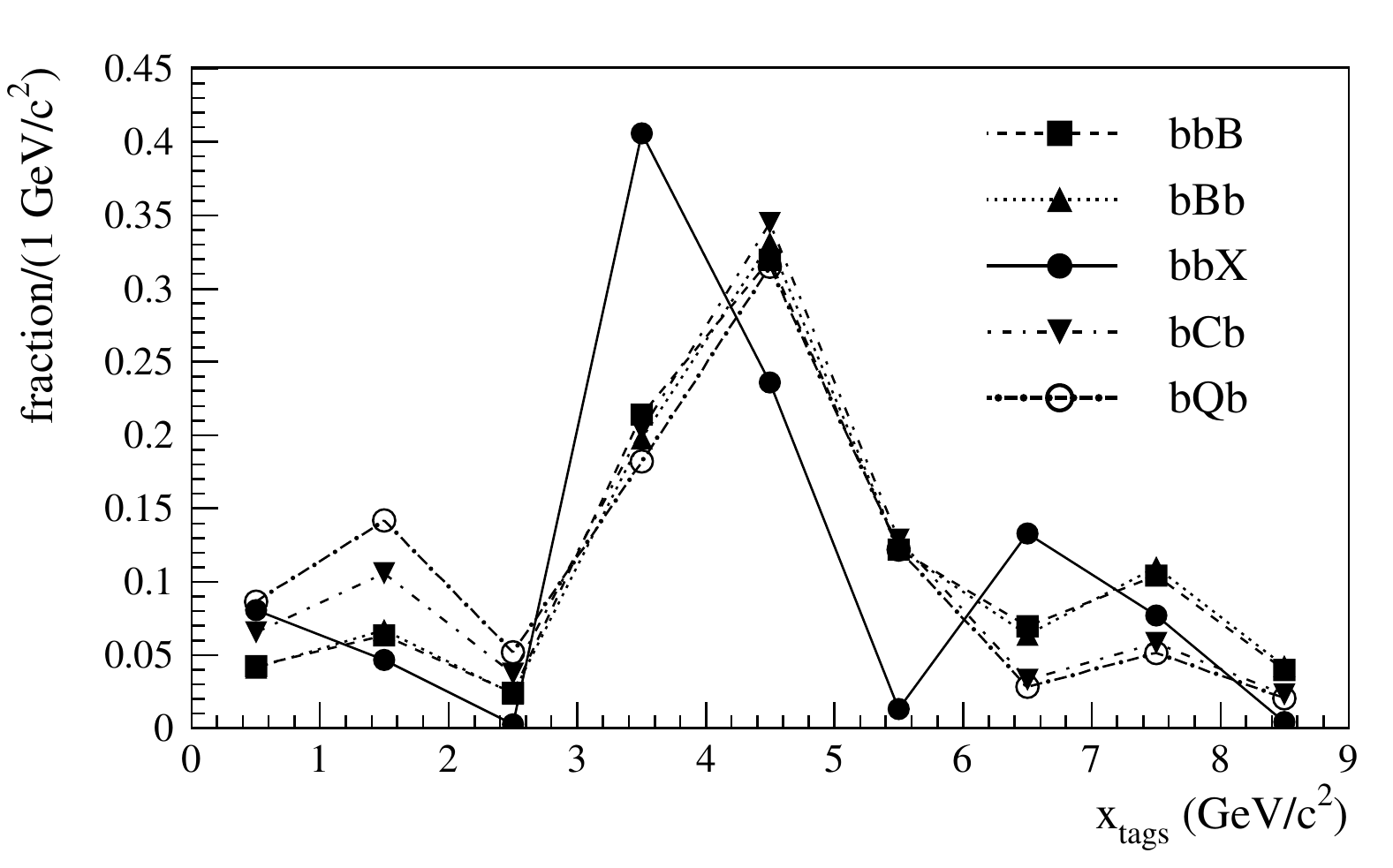}
\end{center}
\end{figure}

Distributions of $x_{tags}$ for all of the background components are
shown in Fig.~\ref{f:bkgd_shapes_xtags}.  The backgrounds separate
into three groups in this variable, with $bCb$ and $bQb$ more
prominent in the bins of lower $x_{tags}$, $bbB$ and $bBb$ more
prominent in the higher $x_{tags}$ bins, and $bbX$ with a different
shape due to the non-$b$ flavor of the tag in the third leading jet in
those events.  A neutral scalar signal, because it contains three
$b$-jets, would look very similar to the $bbB$ and $bBb$ backgrounds
in $x_{tags}$.

\subsubsection{Background Normalization Predictions}
\label{s:bkgdnorm}

Our background model requires information only on the shapes of the
various templates, with the normalizations determined from a fit to
the data.  However, it is possible to obtain {\em a priori} estimates
of the normalizations, which can be used as starting points for the
fit, using our templates and inputs from the generator-level {\sc
pythia} study discussed at the beginning of this Section.  As
constructed, the templates have total area equal to the number of
$b\bar{b}$+jet events, multiplied by the average efficiency to tag the
additional jet over the entire $b\bar{b}$+jet sample as if it were
always a light-flavor, charm, or bottom jet depending on the assumed
flavor.  All that remains in each case is to multiply by the fraction
of events where the jet truly has the assumed flavor.  For the charm
and bottom cases, the {\sc pythia} study indicates fractions of 4\%
and 2\%, respectively.

For the light-flavor cases $bbQ$ and $bQb$, 
we use the observed numbers of events with
one or more negative tags to estimate the normalizations of these
components by extending the calculation in Eq.~\ref{e:nonbb} to the
case of three tags

\begin{align}
\label{e:bbq_bqb_norm}
N_{bbQ} &= N_{++-} - \lambda N_{+--} + \lambda^2 N_{---} \\
N_{bQb} &= N_{+-+} - \lambda (N_{+--}+N_{--+}) + \lambda^2 N_{---}
\end{align}

\noindent where the $N$ are the numbers of observed events with the
indicated positive/negative tag patterns.  In the case of the two
leading jets containing a positive and a negative tag, for example
$N_{+-+}$, the negative tag can be in either the first or second
leading jet.  The factor $\lambda$ is the same fake tag asymmetry
factor used in Eq.~\ref{e:nonbb}.

We emphasize that these estimates are never used as constraints in any
fits; the normalizations of the background components are always
derived strictly from the data sample itself without any theoretical input on
jet flavor fractions.  We will however use these {\em a priori}
estimates as starting points in Sec.~\ref{s:results} for
estimating the sensitivity of our search.

\subsubsection{Background-Only Fit to the Data} 

We fit the background and signal templates to the data using a binned
maximum-likelihood fit.  The likelihood function is a joint
probability of the Poisson likelihood for each bin $\nu_{ij}^{n_{ij}}
e^{-\nu_{ij}}/n_{ij}!$, where $n_{ij}$ is the number of observed events
in the $i$-th bin of $m_{12}$ and the $j$-th bin of $x_{tags}$, and
the expectation in that bin $\nu_{ij}$ is given by

\begin{equation}
\nu_{ij} = \sum_b N_b f_{b,ij} + N_s f_{s,ij}
\end{equation}

\noindent where $b$ represents the five background templates, $f_{b,ij}$
and $f_{s,ij}$ are the bin contents of the various backgrounds ($f_b$)
and of the neutral scalar signal ($f_s$), and the five $N_b$ and optionally $N_s$
are the free parameters of the fit which represent the normalizations
of each component.  We normalize all background and signal
templates to unit area when performing this fit, so that the $N_b$ and
$N_s$ parameters will correspond to the numbers of events in the sample
assigned to each template.

\begin{figure}
\caption{Fit of the triple-tagged data sample using only the QCD 
background templates, in the $m_{12}$ (a) and $x_{tags}$ (b)
projections.  The differences between the data and the fit model are
shown in the lower section of each figure.}
\label{f:fit_data_bkgd}
\begin{center}
\includegraphics[width=\figwidth]{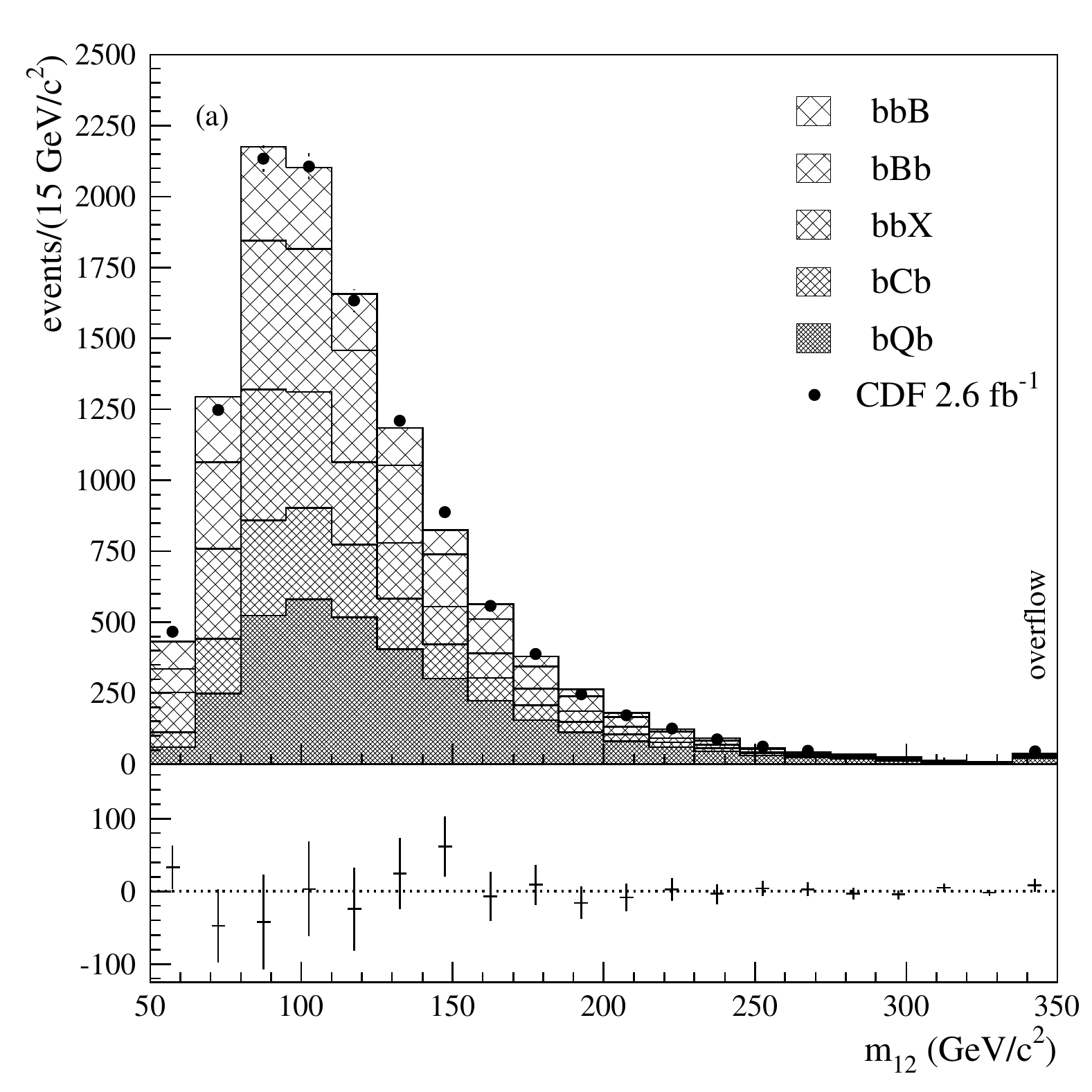}
\includegraphics[width=\figwidth]{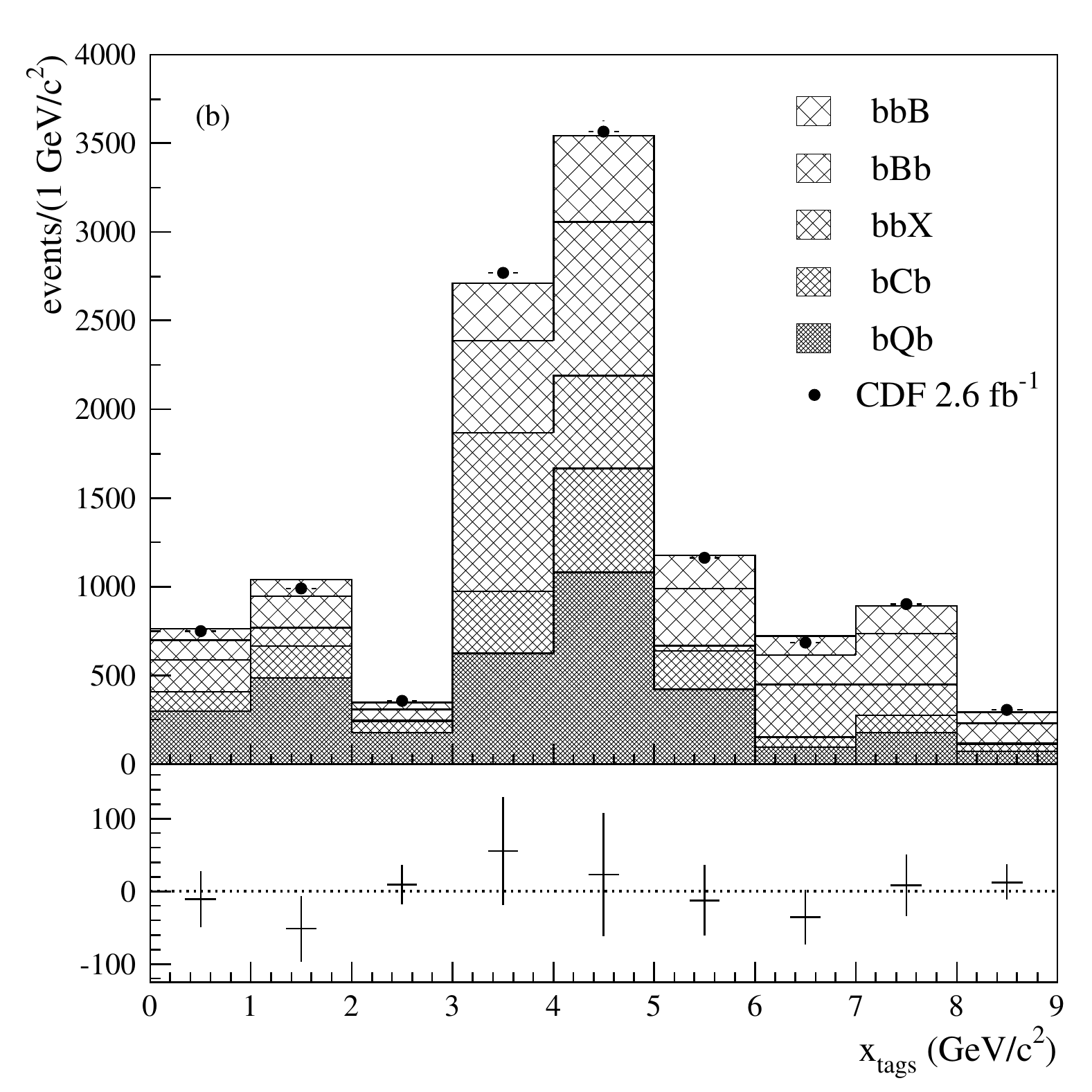}
\end{center}
\end{figure}

Fig.~\ref{f:fit_data_bkgd} shows the result of a fit of the 11 490
triple-tagged events observed in the data using only the background
templates ($N_s$ fixed to zero) and with no systematic errors.  Only
the projections onto each axis are shown for clarity.  The post-fit
$\chi^2/dof$ between the observed data and best-fit background is
185.8/163~=~1.140.  The numbers of fitted events for each background
type are given in Table~\ref{t:fit_data_bkgd} and compared to the
predictions derived from the {\sc pythia} jet flavor fractions.  Good
agreement is observed for all background components.  This comparison
does not demonstate the ability of {\sc pythia} to correctly model the
$m_{12}$ spectrum observed in the data, it tests only the overall
numbers of events predicted for each flavor composition but not their
kinematics.  The good agreement between the fitted number of
$bQb$ events and the data-driven prediction of the normalization does
indicate that we are not missing any sizeable background component
with fewer than two $b$-jets, because that component would be expected
to show up at higher $m_{12}$ values due to the bias towards
high-$E_T$ jets produced by fake tags.

\begin{table}
\caption{Numbers of fitted events for each background type, compared
to the estimates derived from the {\sc pythia} heavy flavor fractions.}
\label{t:fit_data_bkgd}
\begin{center}
\begin{ruledtabular}
\begin{tabular}{ccc}
component & estimate & $N_{fit}$ \\
\hline
$bbB$ & 1300 & 1520$\pm$540 \\
$bBb$ & 2950 & 2620$\pm$550 \\
$bbX = bbQ + bbC$ & 1350+640 = 1990 & 2210$\pm$160 \\
$bCb$ & 1380 & 1710$\pm$630 \\
$bQb$ & 3480 & 3430$\pm$390 
\end{tabular}
\end{ruledtabular}
\end{center}
\end{table}

In order to fully judge the quality of the background-only fit
and whether it adequately describes the data, we require a framework
that allows for the introduction of systematic uncertainties.  We also
need to be able to calculate the significance of any possible signal
contribution after accounting for systematic uncertainties.  The
procedure we adopt uses ensembles of simulated experiments, where the
simulated experiments include the effects of systematic uncertainties
and the fitting procedure is the same as described above.  We describe
the systematic uncertainties which we consider in the next Section and
the simulated experiments procedure in Sec.~\ref{s:results}.

\section{Systematic Uncertainties}
\label{s:systematics}

Several sources of systematic uncertainty on the signal and background
contributions are considered.  A summary is shown in
Table~\ref{t:systematics}.  Modeling uncertainties can affect both the
normalization of the fit templates (denoted `rate' in the Table) and
the distributions of $m_{12}$ and $x_{tags}$ (denoted `shape').  Shape
uncertainties are introduced by modifying the templates using an
interpolation procedure~\cite{p:histinterp}.

\begin{table*}
\caption{Summary of systematic uncertainties.}
\label{t:systematics}
\begin{center}
\begin{ruledtabular}
\begin{tabular}{cccc}
source & variation & applies to & type \\
\hline
luminosity & $\pm$6\% & signal & rate \\
Monte Carlo statistics & $\pm$2\% & signal & rate \\
selection efficiency & $\pm$5\% per jet & signal & rate \\
PDFs & $^{+3.5}_{-4.5}$\% & signal & rate \\
jet energy scale & $\pm$4.5\% & signal & rate/shape \\
$b/c$ $m_{tag}$ & 3\% & signal/backgrounds & shape \\
mistag $m_{tag}$ & 3\% & backgrounds & shape \\
mistag asymmetry factor $\lambda$ & 1.4 $\pm$ 0.2 & backgrounds & rate/shape \\
heavy flavor fractions & $\pm$50\% & backgrounds & rate
\end{tabular}
\end{ruledtabular}
\end{center}
\end{table*}

Rate uncertainties on the signal contribution relate to the number of
signal events expected for a given cross section.  They include the
integrated luminosity of the data sample, the statistical errors due
to the finite size of the simulated signal samples, the efficiency of
the trigger and SECVTX tagging requirements, and the effect on the
efficiency due to uncertainties on parton distribution functions
(PDFs).  For the PDF uncertainty we apply the twenty eigenvector
variations of the CTEQ 6.5M set.

Modeling of the energy scale of jets introduces uncertainties both on
the acceptance for signal events to pass the event selection and on the
$m_{12}$ spectrum of these events.  No energy scale modeling
uncertainty is assigned to the background templates since they are
derived from the data.

The $x_{tags}$ variable introduces an uncertainty due to modeling of
the $m_{tag}$ spectrum of the SECVTX displaced vertices.  This
uncertainty affects only the shape of the $x_{tags}$ distribution and
has no effect on the estimated signal acceptance.  For the simulated signal events, all three SECVTX vertex masses are varied, while for the
backgrounds only the mass of the simulated third tag in the event is
varied because the other two tag masses in each event come directly
from the data.

Varying the value of $\lambda$ used to subtract the non-$b\bar{b}$
component from the double-tagged events changes the shapes of the
resulting corrected background templates, and also the predicted
normalizations of the $bbQ$ and $bQb$ templates.

We assign 50\% uncertainty to the 2\% ($b$) and 4\% ($c$) jet flavor
fractions from {\sc pythia} which are used to obtain the {\em a
priori} normalization estimates of the background components.  This
variation is used only when throwing the simulated experiments used to
estimate the sensitivity.  It is not used to constrain any of the
templates in the fits.  The results are largely insensitive to the
size of this variation, so long as it is large relative to the
precisions obtained on each template from the fit but not so large
that it causes the simulated experiments to often have zero
contribution from any of the background components.  When performing
the variation we assume that $bbB$ and $bbC$ are 100\% correlated
because they are likely to involve the same underlying physics
processes; the same holds for $bBb$ and $bCb$.  No correlation is
assumed between $bbB$ and $bBb$ or $bbC$ and $bCb$.

\section{Resonance Search in the Triple-Tagged Data}
\label{s:results}

We perform fits of the data using the background templates and
templates for a neutral scalar in the mass range of 90-350 GeV/$c^2$.
These fits are identical to the one shown in
Fig.~\ref{f:fit_data_bkgd} except that in addition to floating the
background normalizations we also release the constraint on the
template representing a possible resonant component of the data.  We
use a modified frequentist $CL_S$ method~\cite{p:cls} to compute the
sensitivity and set 95\% confidence level upper limits on the cross
section for production of a narrow scalar as a function of mass.  We
compare the data to the best-fit background plus signal model for the
mass point with the most significant excess.  Finally, we interpret
our results as limits on $\tan\beta$ in the MSSM as a function of the
pseudoscalar Higgs boson mass $m_A$, including the effects of the
Higgs boson width.

\subsection{Cross section times branching ratio limits}

The limit calculations are performed using a custom program based on
the {\sc mclimit} package \cite{p:mclimit}.  It performs the fitting
of the background and signal templates using either the observed data
or simulated experiments, and calculates confidence levels using the
$CL_s$ method.  The test statistic employed is the difference in
$\chi^2$ between fits using only the background templates and fits
using both background and signal templates.

\subsection{Simulated Experiments}

Simulated experiments are generated based on the background
predictions in Table~\ref{t:fit_data_bkgd}.  The number of signal
events generated depends on the assumed $\sigma\times BR$, the
integrated luminosity, and the acceptance shown in
Fig.~\ref{f:acc_vs_mh}.  The predictions for the numbers of each
background type and for the signal are randomly varied for each
simulated experiment according to the systematic uncertainties shown in
Table~\ref{t:systematics}.  The distributions of $m_{12}$ and
$x_{tags}$ are also randomly varied using histogram interpolation.
The resulting background and signal templates are summed to obtain
estimates for the number of events in each bin of $m_{12}$ and
$x_{tags}$.  These are input to a Poisson random-number generator to
produce integer bin counts for the simulated experiment with the
appropriate statistical variations.  These are fit using the
default background and signal templates to build probability
densities of the test statistic for various values of $\sigma\times
BR$.  The fits of either the observed data or simulated experiments
always use the unmodified templates.  The systematic uncertainties are
only applied when building the simulated experiments.

\subsection{Limit Results}

The median expected limits on $\sigma\times BR$ for statistical errors only
and with full systematic uncertainties applied are shown in
Table~\ref{t:xs-limits}, along with the observed limits.  The
systematic uncertainties increase the limits by 15-25\% relative to the
no-systematics case.

\begin{table}[tp]
\caption{Median expected and observed limits on
$\sigma(p\bar{p}\rightarrow b\phi)\times \mathcal{B}(\phi\rightarrow b\bar{b})$, in pb.}
\label{t:xs-limits}
\begin{center}
\begin{ruledtabular}
\begin{tabular}{cccc}
$m_\phi$ & no systematics & full systematics & observed \\
\hline
  90 & 39.8 & 48.8 & 26.4 \\
 100 & 41.3 & 50.8 & 32.6 \\
 110 & 22.7 & 28.0 & 27.8 \\
 120 & 20.1 & 23.0 & 34.5 \\
 130 & 13.4 & 15.5 & 28.8 \\
 140 & 12.0 & 13.8 & 33.8 \\
 150 &  9.2 & 10.7 & 28.0 \\
 160 &  8.1 &  9.1 & 22.2 \\
 170 &  6.3 &  7.3 & 16.7 \\
 180 &  6.0 &  6.7 & 11.6 \\
 190 &  5.2 &  6.1 &  7.7 \\
 200 &  4.9 &  5.5 &  6.4 \\
 210 &  4.3 &  4.9 &  5.1 \\ 
 220 &  4.1 &  4.6 &  5.0 \\
 230 &  3.6 &  4.2 &  4.8 \\
 240 &  3.5 &  4.1 &  5.1 \\
 250 &  3.2 &  3.9 &  4.9 \\
 260 &  3.1 &  3.7 &  4.9 \\
 270 &  2.9 &  3.5 &  4.7 \\
 280 &  2.9 &  3.4 &  4.5 \\
 290 &  2.7 &  3.3 &  4.4 \\
 300 &  2.7 &  3.2 &  4.3 \\
 310 &  2.5 &  3.3 &  4.9 \\
 320 &  2.5 &  3.1 &  4.7 \\
 330 &  2.7 &  3.1 &  4.8 \\
 340 &  2.5 &  3.0 &  4.8 \\
 350 &  2.5 &  3.3 &  5.6
\end{tabular}
\end{ruledtabular}
\end{center}
\end{table}

\begin{figure}[tp]
\caption{Median, 1$\sigma$, and 2$\sigma$ expected limits, and 
observed limits on narrow resonance production versus $m_\phi$ on linear
(a) and logarithmic (b) scales.}
\label{f:xs-limits}
\begin{center}
\includegraphics[width=\figwidth]{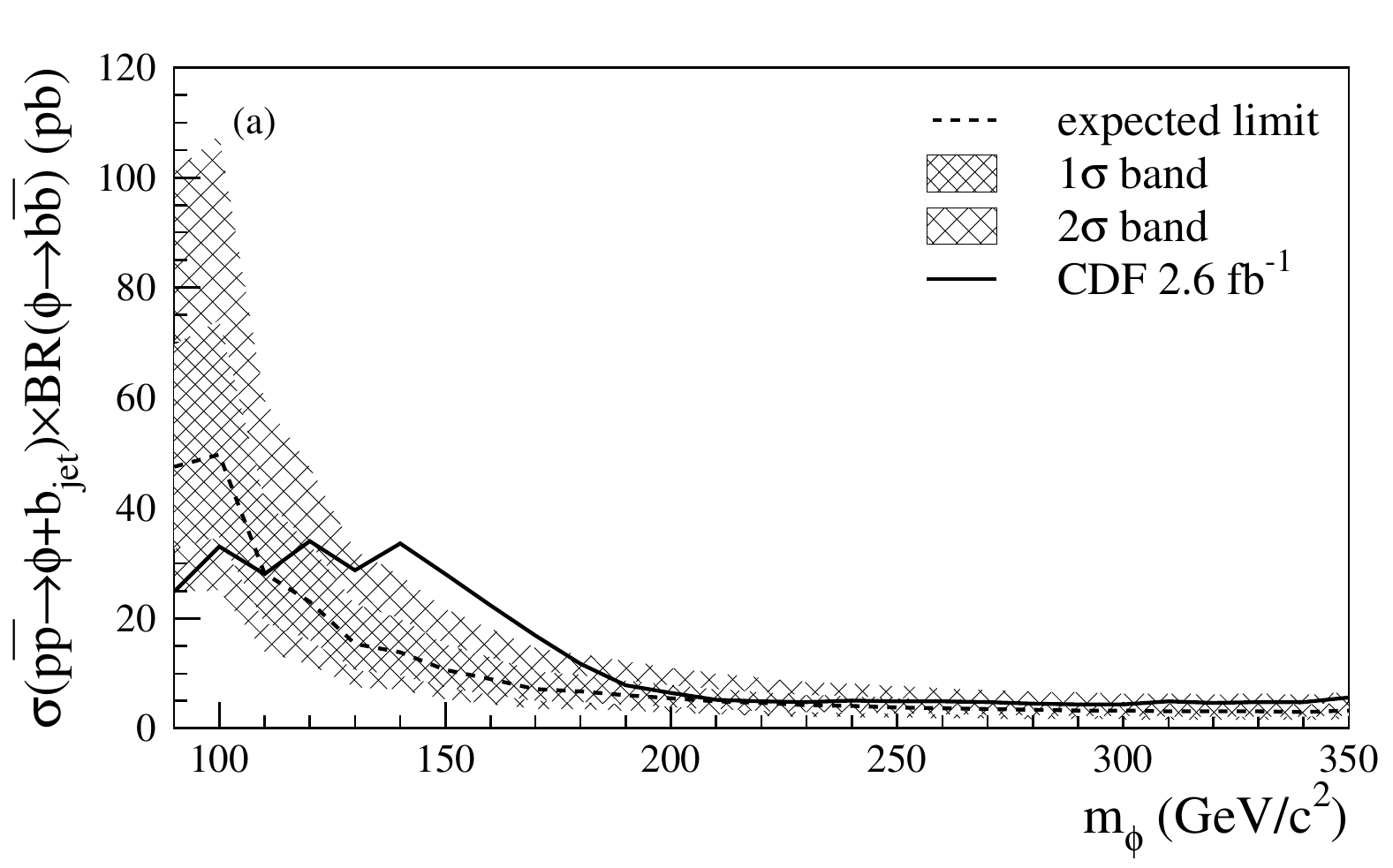}
\includegraphics[width=\figwidth]{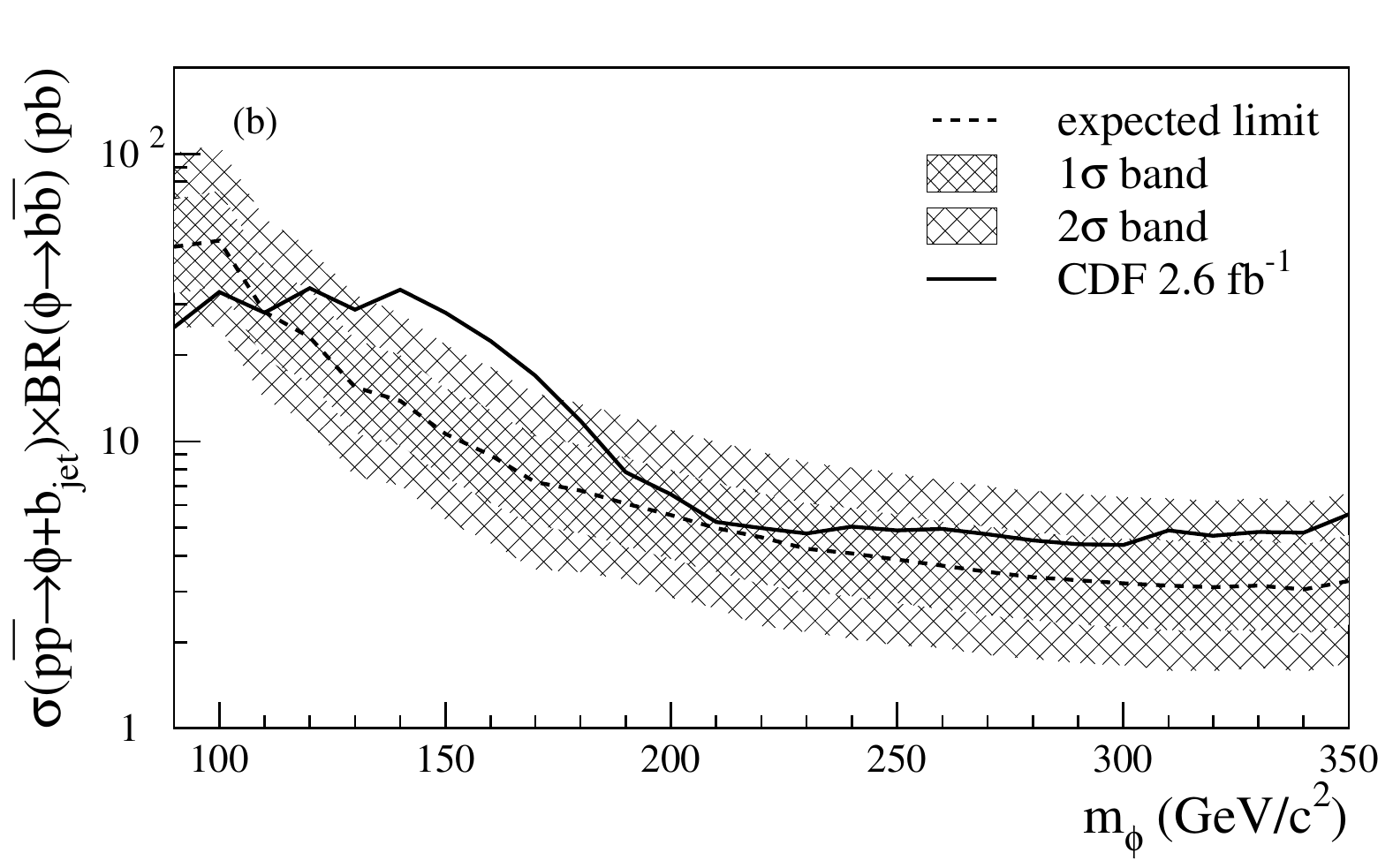}
\end{center}
\end{figure}

The expected and observed limits for the full systematics case are
plotted as a function of the narrow scalar mass in
Fig.~\ref{f:xs-limits}.  Also shown are the bands resulting from
calculating the expected limits using the $\pm1\sigma$ and
$\pm2\sigma$ values of the test statistic from simulated experiments
containing no signal.  We observe a positive deviation of greater than
2$\sigma$ from the expectation in the mass region of 130-160
GeV/$c^2$.  The most significant discrepancy is at $m_\phi =
150$~GeV/$c^2$, with a 1-$CL_b$ $p$-value of 0.23\%.  Including the
trials factor to account for the number of mass points searched, we
expect to see a deviation of this magnitude at any mass in the range
which we test (90-350 GeV/$c^2$ in steps of 10 GeV/$c^2$) in 2.5\% of
background-only pseudoexperiments.

\begin{figure}
\caption{Fit of the triple-tagged data sample using the QCD background 
templates and the signal template for $m_\phi = 150$ GeV/$c^2$, in the
$m_{12}$ (a) and $x_{tags}$ (b) projections.  The differences between
the data and the fit model are shown in the lower section of each
figure.}
\label{f:fit_data_mh150}
\begin{center}
\includegraphics[width=\figwidth]{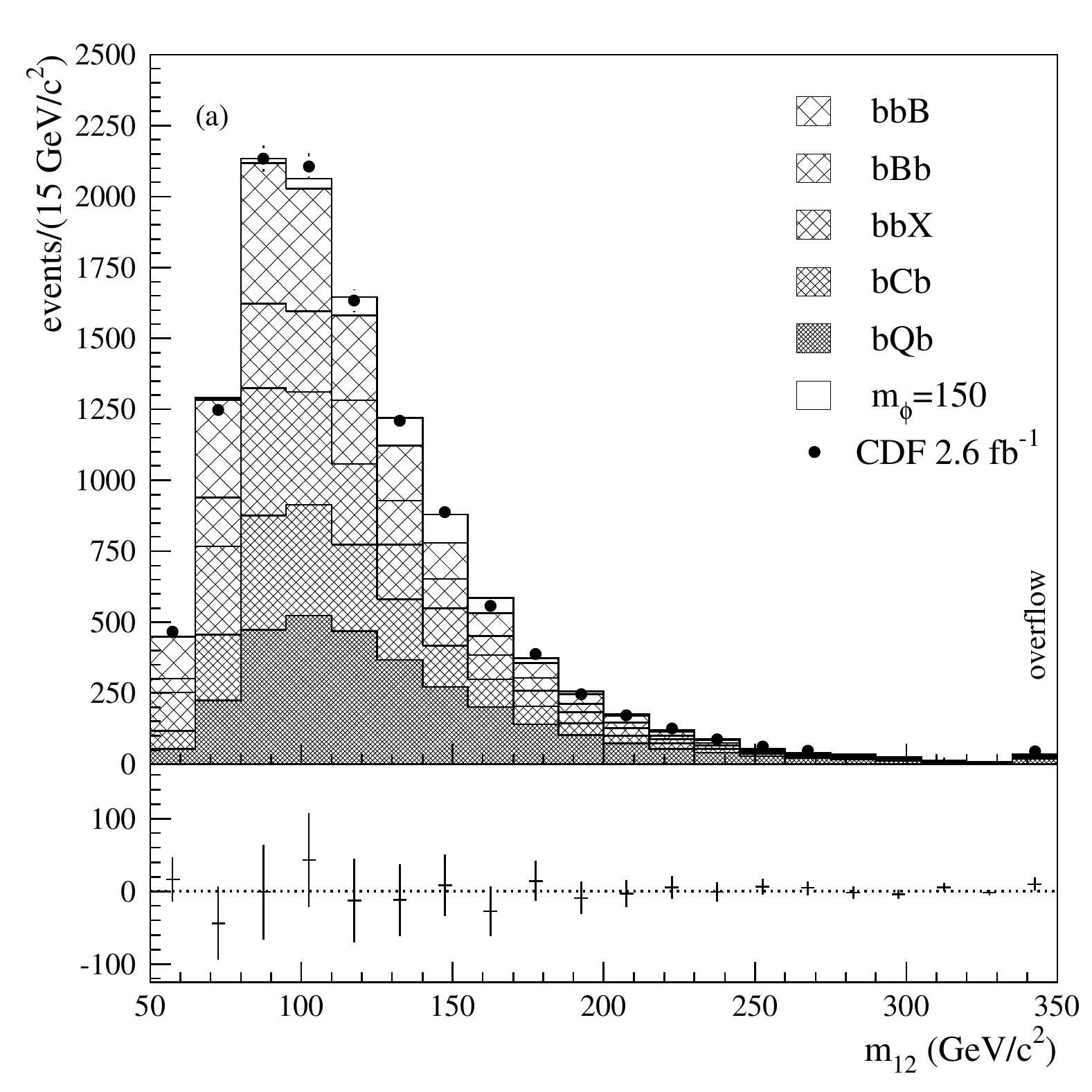}
\includegraphics[width=\figwidth]{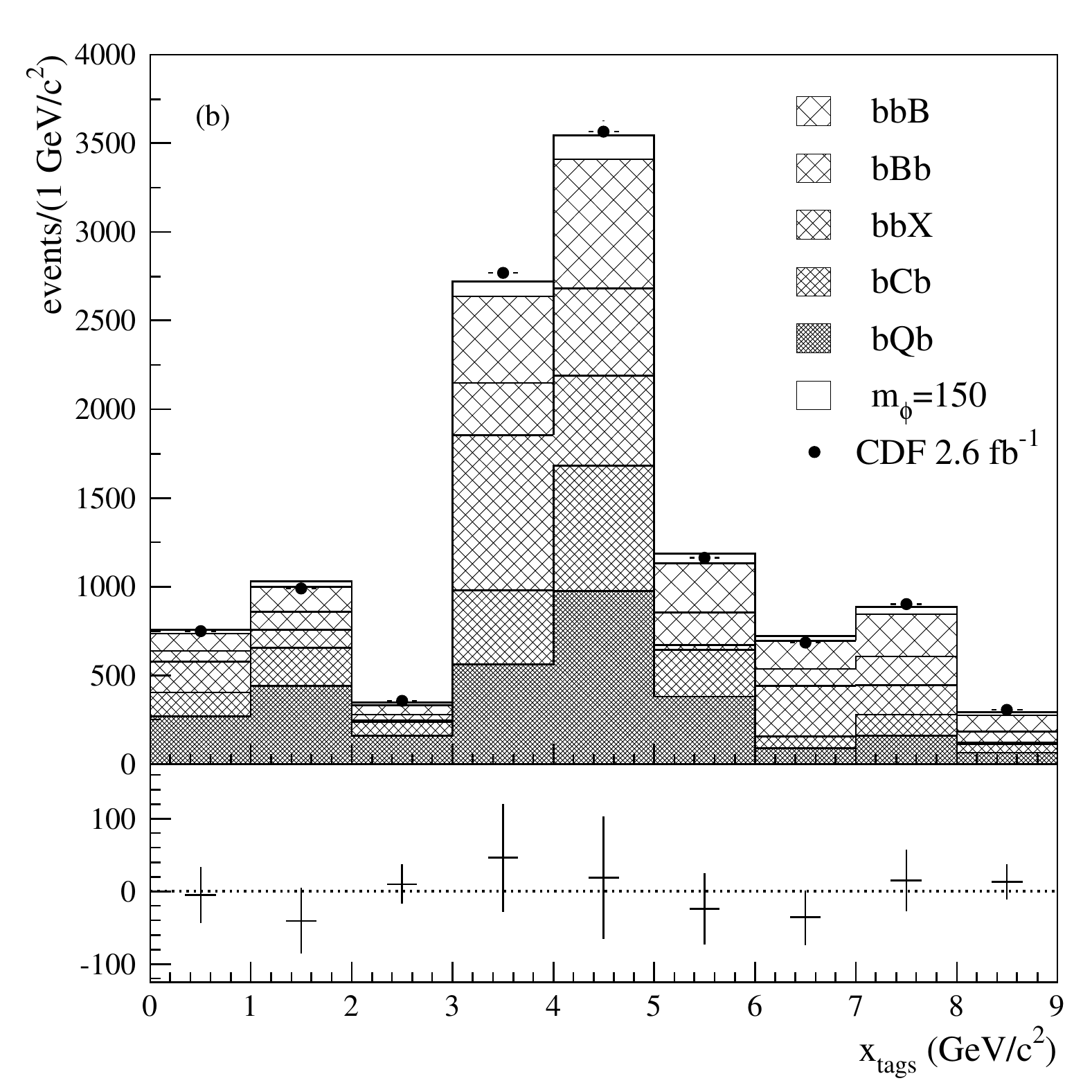}
\end{center}
\end{figure}

The results of the fit of the observed data for a narrow scalar mass of
150~GeV/$c^2$ are shown in Fig.~\ref{f:fit_data_mh150} and
Table~\ref{t:fit_data_mh150}.  In this case the $\chi^2/dof$ is
171.2/162~=~1.057, with the fit assigning 420$\pm$130 events to the
signal template.  If interpreted as narrow scalar production
this corresponds to a cross section times branching ratio of about
15~pb within our Higgs-like production model.

\begin{table}
\caption{Numbers of fitted events for each background type and for
a narrow scalar signal with $m_\phi = 150$ GeV/$c^2$.}
\label{t:fit_data_mh150}
\begin{center}
\begin{ruledtabular}
\begin{tabular}{cc}
component & $N_{fit}$ \\
\hline
$bbB$ & 2280$\pm$600 \\
$bBb$ & 1490$\pm$670 \\
$bbX$ & 2150$\pm$160 \\
$bCb$ & 2050$\pm$630 \\
$bQb$ & 3100$\pm$400 \\
Higgs & 420$\pm$130
\end{tabular}
\end{ruledtabular}
\end{center}
\end{table}

\subsection{Checks of the Background Model}

Several checks are made to investigate if the slight excess in the
140-170~GeV/$c^2$ mass region might be due to a neglected
background contribution or mismodeling of one or more of the
background templates.

One possible explanation is the effect of neglecting the component of
the multijet background with fewer than two $b$-jets.  The components
with at least two charm jets are found to be accommodated by residual
$c\bar{c}$ contributions in the double-tagged sample used to construct
the background estimates.  To check the effect of backgrounds with at
least two falsely-tagged light-flavor jets, we introduce a template
into the fit derived from events with one positive and two negative
tags.  We find the fit prefers to assign $\sim$1\% of the sample to
this template, with slightly reduced fit quality as determined from
the $\chi^2/dof$ (170.5/161 = 1.059).  The change in the fitted excess
is positive and less than 5\%.

We return to the question of $t\bar{t}$ pair production and
$Z\rightarrow b\bar{b}$ + jets backgrounds, which are neglected in the
fit.  We expect around 30 and 100 events from these sources,
respectively.  The $Z\rightarrow b\bar{b}$ + jets background would not
need to be explicitly included in the fit even if it were much larger,
because it is already represented in the double-tagged events used to
construct the background templates.  The jets which accompany
$Z\rightarrow b\bar{b}$ are similar to the jets in multijet $b\bar{b}$
+ jets events, so the fraction of $Z\rightarrow b\bar{b}$ + jets in
the double-tagged background sample is correctly translated into the
correct fraction to account for the $Z\rightarrow b\bar{b}$
contribution to the triple-tagged signal sample.  The $t\bar{t}$
contribution is also partially accounted for by this mechanism,
although the jet flavor composition is not as similar due to enhanced
charm production from $W\rightarrow c\bar{s}$ decays.  Due to the
smallness of the overall $t\bar{t}$ contribution the remaining
contribution can safely be neglected.

Mismodeling of the instrumental bias introduced by simulating the
effect of the third tag could distort the backgound templates and
produce an apparent excess.  We test our sensitivity to this effect by
replacing the $E_T$-dependence of the SECVTX tag efficiency for
$b$-jets which we measured in the data with that predicted by our full
detector simulation.  This change is much larger than the precision
with which we measure the $E_T$-dependence in the data.  Fitting the
data with these modified templates, we find changes to the
normalizations of the individual background templates of 50-100 events and
virtually no change in the summed best-fit background model.  We perform a similar test on the $E_T$-dependence of the
false tag rate used to construct the $bbQ$ and $bQb$ templates, in
this case replacing the $E_T$-dependence from the full detector
simulation with an estimate derived from negative tags in the data.
Fitting with these modified templates we find changes in the
background normalizations consistent with statistical fluctuations,
and again little change in the total background model.

\subsection{MSSM interpretation}

To interpret the data in MSSM scenarios, we must know the
production cross section for Higgs boson events with a given pseudoscalar
mass $m_A$ as a function of $\tan\beta$.  At tree level this can be
computed~\cite{p:scenarios} as

\begin{equation}
\sigma_{MSSM} = 2\times \sigma_{SM}\times \tan^2\beta \times 0.9
\end{equation}

\noindent where $\sigma_{SM}$ is the standard model cross section for
a Higgs boson of mass $m_A$, the factor of two reflects the degeneracy
between $A$ and $h/H$, and 0.9 is the branching ratio
$\mathcal{B}(A\rightarrow b\bar{b})$.

In order to go beyond tree level, we must consider the
effects of loop corrections which can enhance the cross
section by more or less than $\tan^2\beta$ depending upon the MSSM
scenario.  We must also include the effects of the Higgs boson width
which can become significant when the down-type couplings are enhanced
by such large factors.  This means that not only the amount of signal
expected but also the properties of that signal such as the
reconstructed $m_{12}$ spectrum will change depending upon the value
of $\tan\beta$ in the scenario under consideration.

In Refs.~\cite{p:scenarios,p:balazs} an approximate expression for the cross
section times branching ratio for Higgs boson production in the MSSM,
including loop effects, is given as:

\begin{equation}
\label{e:scenario}
\sigma(b\bar{b}\phi)\times \mathcal{B}(A\rightarrow b\bar{b}) \simeq
2\sigma(b\bar{b}\phi)_{SM}\frac{\tan^2\beta}{(1+\Delta_b)^2}\times
\frac{9}{(1+\Delta_b)^2+9}
\end{equation}

\noindent where $\phi$ is a Higgs boson (either the SM variety or one of
$h/H/A$), $\sigma(b\bar{b}\phi)_{SM}$ is the SM cross section, the
factor of two comes from the degeneracy of $A$ with either $h$ or $H$,
and the loop effects are incorporated into the $\Delta_b$ parameter.
For our purposes it is important only to note that $\Delta_b$ is
proportional to the product of $\tan\beta$ and the Higgsino mass
parameter $\mu$.  Sample values of $\Delta_b$ given in
Ref.~\cite{p:scenarios} are -0.21 for the $m_h^{max}$ scenario and
-0.1 for the no-mixing scenario (at $\mu = -200$ GeV and $\tan\beta =
50$).  It is apparent that negative values of $\mu$ and hence of
$\Delta_b$ will increase the MSSM Higgs boson yield at a given $\tan\beta$
above the tree level values and result in stronger limits on
$\tan\beta$, while scenarios with $\mu$ positive will produce the
opposite effect.  Using Eq.~\ref{e:scenario} we can predict the Higgs
boson yield for any value of $\tan\beta$ and $\Delta_b$ and therefore derive
limits in any desired scenario.

The limits shown in Fig.~\ref{f:xs-limits} apply only to narrow
scalars such as the standard model Higgs boson.  If the cross section is
increased by scaling the $b\bar{b}\phi$ coupling, as happens in the MSSM,
then the width of the Higgs boson will increase as well.  In order to
account for this we convolute the cross section shown in
Fig.~\ref{f:mcfm-xs} with a relativisitic Breit-Wigner to produce
cross section lineshapes for various values of the Higgs boson pole mass,
$\tan\beta$, and $\Delta_b$.  Parametrizations of the partial widths
$\Gamma_{b\bar{b}}$ and $\Gamma_{\tau\tau}$ as functions of $m_A$ and
$\tan\beta$ are obtained from the {\sc
feynhiggs}~\cite{p:feynhiggs1,*p:feynhiggs2,*p:feynhiggs3,*p:feynhiggs4}
program, with $\Gamma_{b\bar{b}}$ also dependent on $\Delta_b$.

Changing the width of the Higgs boson also changes the total cross
section as a function of the pole mass.  We integrate the broadened
cross section described above for $m_\phi > 50$ GeV/$c^2$ (where the
acceptance for a narrow Higgs drops to zero) and divide by the cross
section value expected for a narrow Higgs to derive a correction
factor.  This factor ranges from 1.0-0.8 for pole mass of 90 GeV/$c^2$
to 1.0-1.1 for 180 GeV/$c^2$, for $\tan\beta$ from 40-120.  The factor
drops below 1 for low pole masses because part of the broadened cross
section falls below the cutoff at 50 GeV/$c^2$.  This information is
needed when computing the expected number of events for a given Higgs
boson mass and $\tan\beta$ value in the limits calculator.

\begin{figure}[tp]
\caption{Median, 1$\sigma$, and 2$\sigma$ expected limits, and the
observed limits versus $m_A$, including the Higgs boson width and for
$\Delta_b = 0$.}
\label{f:tb-limits-wide}
\begin{center}
\includegraphics[width=\figwidth]{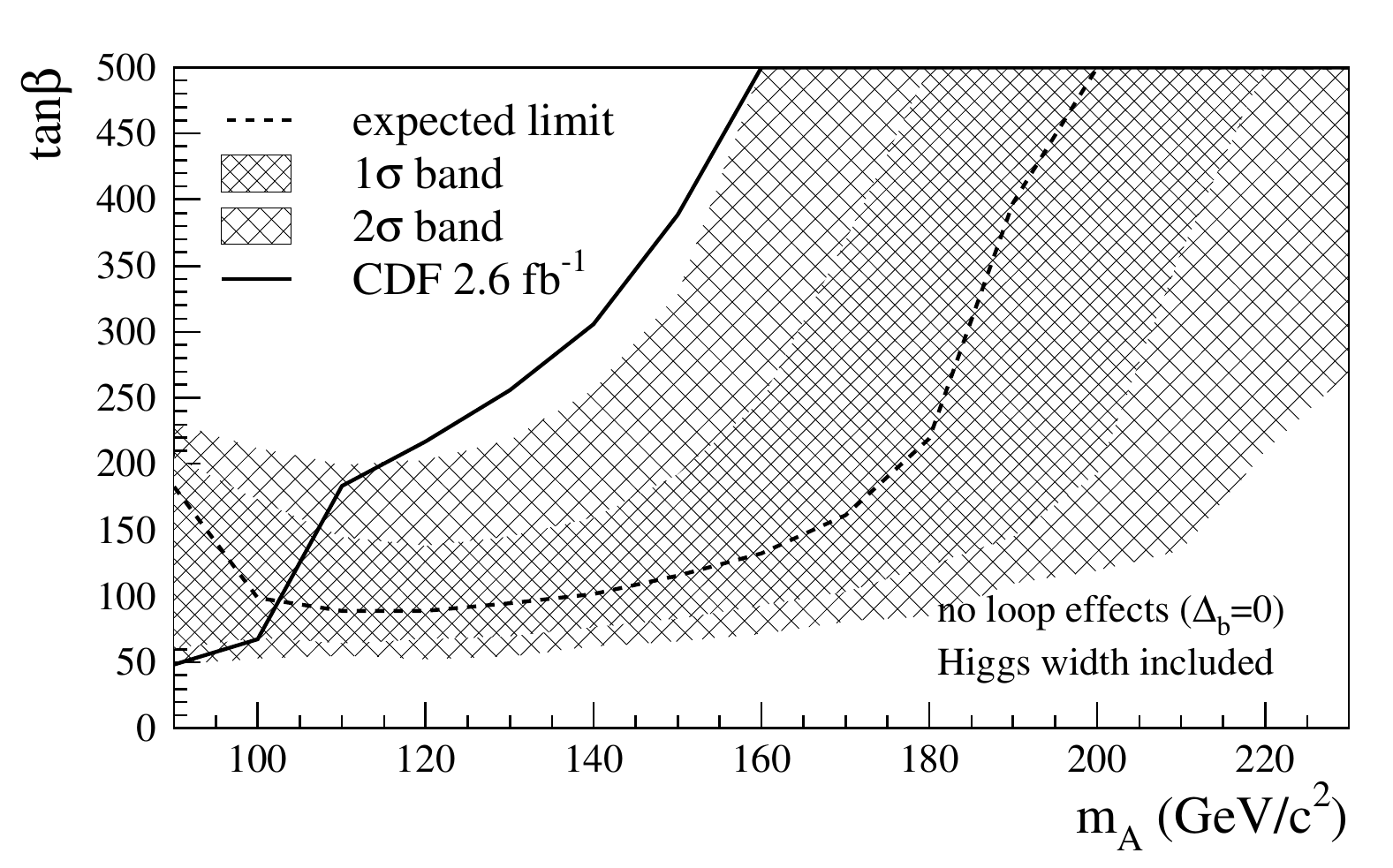}
\end{center}
\end{figure}

Fit templates for the Higgs boson signal as a function of $\tan\beta$ are
constructed by combining the narrow-width templates, weighted by the
lineshapes and by the acceptance parametrization shown in
Fig.~\ref{f:acc_vs_mh}.  We scan over $\tan\beta$ in steps of 5 and
calculate $CL_s$ at each point, and exclude regions with $CL_s >
0.05$.  The limits obtained are shown in Fig.~\ref{f:tb-limits-wide}
for $\Delta_b = 0$.  The sensitivity begins to degrade rapidly for
Higgs boson masses above 180~GeV/$c^2$, where the values of $\tan\beta$
required to produce an observable cross section result in an $m_{12}$
spectrum that no longer displays a mass peak due to the large width of
the Higgs boson.

\begin{figure}[tp]
\caption{Median, 1$\sigma$, and 2$\sigma$ expected limits, and the
observed limits versus $m_A$, including the Higgs boson width, for the
$m_h^{max}$ scenario with $\mu = -200$ GeV.}
\label{f:tb-limits-mhmax}
\begin{center}
\includegraphics[width=\figwidth]{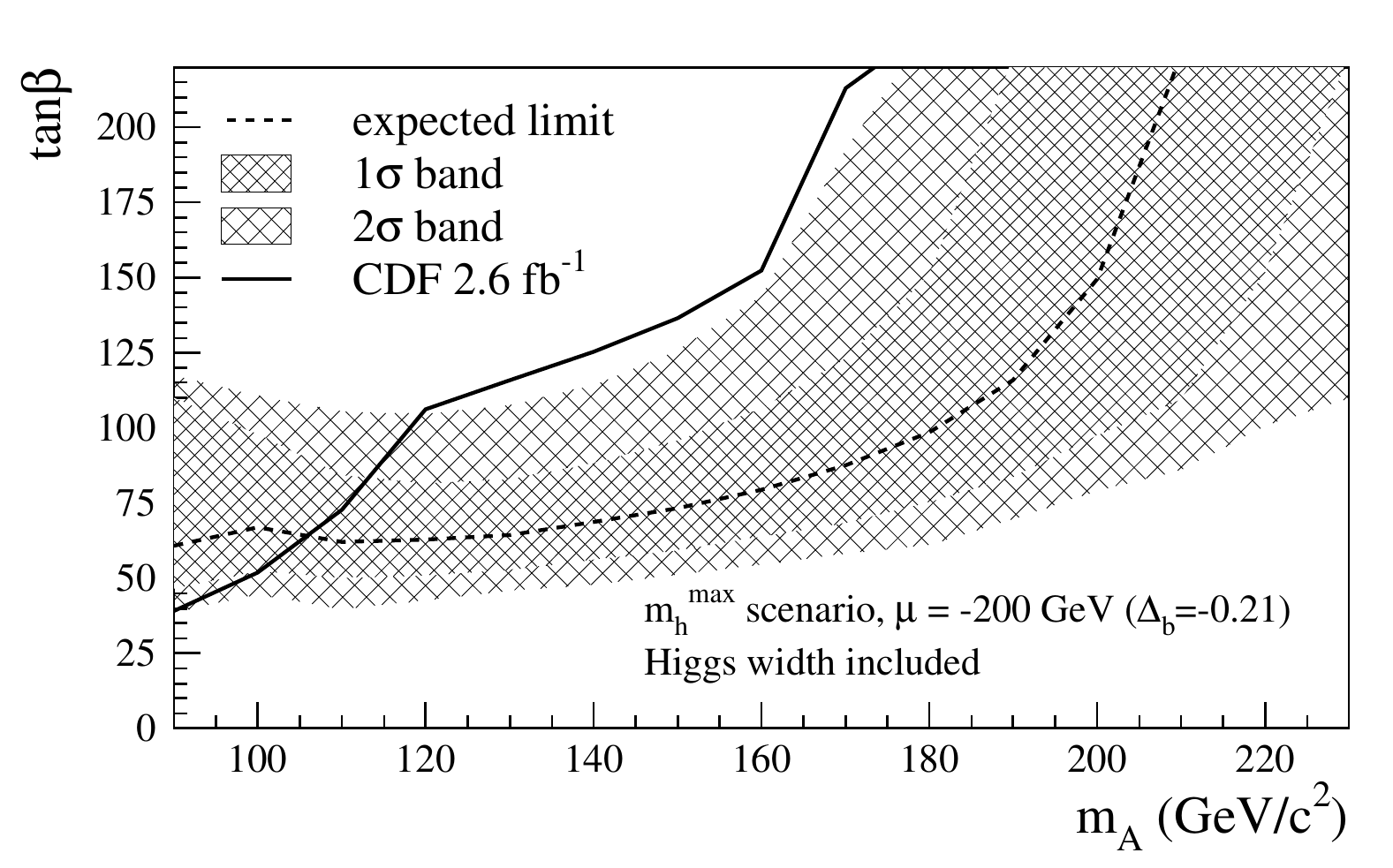}
\end{center}
\end{figure}

Along with the $\Delta_b=0$ case, limits are also generated for the
$m_h^{max}$ scenario with $\mu = -200$ GeV and are shown in
Fig.~\ref{f:tb-limits-mhmax}.  Because of the relatively large and
negative values of $\Delta_b$ in this scenario, the $\tan\beta$ limits
are much stronger because we expect many more signal events for a
given $\tan\beta$ relative to the $\Delta_b=0$ case.  In both cases
the observed limits in the mass range 120-170~GeV/$c^2$ are slightly
above the $2\sigma$ band, due to the excess of data over the
background model in this region.

\section{Conclusion}
\label{s:conclusion}

A search for resonances produced in association with $b$-quarks is
performed in triple-$b$-tagged three- or four-jet events, using 2.6
fb$^{-1}$ of $p\bar{p}$ collisions from the Tevatron.  This process
could be present at a measurable rate in supersymmetric models with
high values of $\tan\beta$.  We use the mass of the two leading jets
and jet flavor information from the secondary vertex tags to fit for a
Higgs boson component within the heavy flavor multijet background.

We find the data are consistent with the background model
predictions over the entire mass range investigated.  The
largest deviation is observed in the mass region 140-170~GeV/$c^2$,
where data show an excess over background with a
significance of 0.23\% ($2.8\sigma$) at 150~GeV/$c^2$.
If this excess were to be attributed
to the production of a narrow resonance in association with a $b$-jet
with kinematics characteristic of Higgs boson production, it would
correspond to a production cross section times branching ratio of
about 15~pb.
We estimate the probablility to observe such a deviation at any mass
in the range 90-350~GeV/$c^2$ at 2.5\% (1.9$\sigma$).  Below
140~GeV/$c^2$ and above 170~GeV/$c^2$ the limits are within 2$\sigma$
of expectations.

The D0 experiment published results for a similar search as the one
performed here in Ref.~\cite{p:d04b-5fb}.  That analysis uses a
multivariate selection and discrimination procedure tuned to the MSSM
Higgs boson hypothesis, whereas here a more general resonance search
is performed.

The data are used to examine two MSSM scenarios.  In the case where
loop effects are small, we find that the growth of the Higgs boson
width as the couplings are enhanced permits only weak limits of
$\tan\beta > 250$ in the mass region around 150~GeV/$c^2$.  In the
$m_h^{max}$ scenario with $\mu$ negative, the enhanced production
through loop effects allows exclusion of $\tan\beta$ values greater
than 40 for $m_A=90$~GeV/$c^2$ and about 90-140 for the mass range
110-170 GeV/$c^2$.  The results in Ref.~\cite{p:d04b-5fb} exclude
values of $\tan\beta$ in the same $m_h^{max}$ with $\mu$ negative
scenario considered here above 50-60 over this mass range.

The MSSM study allows comparison with the results in the $A\rightarrow
\tau\tau$
channel~\cite{p:cdfditau-18fb,p:d0ditau-1fb,p:d0bditau-3fb,p:cmsditau},
which are much less sensitive to the details of the MSSM scenario.
The $\tau\tau$ analyses exclude values of $\tan\beta$ above 25-35 in
the mass range from 90-200~GeV/$c^2$.  Any interpretation of the
observed excess in the results presented here in terms of MSSM Higgs
boson production would therefore be restricted to scenarios with large
negative values of the Higgsino mass parameter $\mu$, where the event
yield in the $b\bar{b}$ decay mode for a given value of $\tan\beta$ is
enhanced.

\bigskip
\begin{acknowledgments}

We thank the Fermilab staff and the technical staffs of the
participating institutions for their vital contributions. This work
was supported by the U.S. Department of Energy and National Science
Foundation; the Italian Istituto Nazionale di Fisica Nucleare; the
Ministry of Education, Culture, Sports, Science and Technology of
Japan; the Natural Sciences and Engineering Research Council of
Canada; the National Science Council of the Republic of China; the
Swiss National Science Foundation; the A.P. Sloan Foundation; the
Bundesministerium f\"ur Bildung und Forschung, Germany; the Korean
World Class University Program, the National Research Foundation of
Korea; the Science and Technology Facilities Council and the Royal
Society, UK; the Institut National de Physique Nucleaire et Physique
des Particules/CNRS; the Russian Foundation for Basic Research; the
Ministerio de Ciencia e Innovaci\'{o}n, and Programa
Consolider-Ingenio 2010, Spain; the Slovak R\&D Agency; the Academy of
Finland; and the Australian Research Council (ARC).

\end{acknowledgments}

\bibliography{note}

\end{document}